\documentclass[11pt]{article}
\usepackage[centertags]{amsmath}
\usepackage{amsmath}
\usepackage{amsfonts}
\usepackage{amssymb}
\usepackage{graphics}
\usepackage[dvips]{graphicx}
\usepackage{amsthm}
\usepackage{newlfont}

\begin{document}

%---------------------------------------------------------------------
\thispagestyle{empty} \setlength{\baselineskip}{1\baselineskip}
\date{}
\title{On Successive Refinement for the Kaspi/Heegard-Berger Problem}
\author{Alina Maor%\thanks{This
%work is part of A. Maor's Ph.D. dissertation.}
\textrm{ } and Neri
Merhav} \maketitle
\begin{center}
Department of Electrical Engineering \\
Technion -- Israel Institute of Technology \\
Technion City, Haifa 32000, Israel \\
{\tt \{alinam@tx, merhav@ee\}.technion.ac.il}
\end{center}
%--------------------------------------------------------------------
\vspace{1.5\baselineskip}
\setlength{\baselineskip}{1.5\baselineskip}
%-------------------------------------------------------------------

\newtheorem{Theorem}{Theorem}
\newtheorem{Lemma}{Lemma}
\newtheorem{Statement}{Statement}
\newtheorem{Corollary}{Corollary}
\newtheorem{Definition}{Definition}
\newtheorem{Claim}{Claim}

\newcommand{\bx}{\emph{\textbf{x}}}
\newcommand{\by}{\emph{\textbf{y}}}
\newcommand{\bz}{\emph{\textbf{z}}}
\newcommand{\bu}{\emph{\textbf{u}}}
\newcommand{\ba}{\emph{\textbf{a}}}
\newcommand{\bb}{\emph{\textbf{b}}}
\newcommand{\bw}{\emph{\textbf{w}}}
\newcommand{\bv}{\emph{\textbf{v}}}

\newcommand{\bA}{\emph{\textbf{A}}}
\newcommand{\bX}{\emph{\textbf{X}}}
\newcommand{\bY}{\emph{\textbf{Y}}}
\newcommand{\bZ}{\emph{\textbf{Z}}}
\newcommand{\bU}{\emph{\textbf{U}}}
\newcommand{\bB}{\emph{\textbf{B}}}
\newcommand{\bW}{\emph{\textbf{W}}}
\newcommand{\bV}{\emph{\textbf{V}}}
\newcommand{\bD}{\emph{\textbf{D}}}
\newcommand{\bS}{\emph{\textbf{S}}}
\newcommand{\bO}{\emph{\textbf{O}}}

\newcommand{\calX}{\mathcal{X}}
\newcommand{\calY}{\mathcal{Y}}
\newcommand{\calS}{\mathcal{S}}
\newcommand{\calC}{\mathcal{C}}
\newcommand{\calU}{\mathcal{U}}
\newcommand{\calV}{\mathcal{V}}
\newcommand{\calA}{\mathcal{A}}
\newcommand{\calB}{\mathcal{B}}
\newcommand{\calZ}{\mathcal{Z}}
\newcommand{\calD}{\mathcal{D}}
\newcommand{\calR}{\mathcal{R}}
\newcommand{\calO}{\mathcal{O}}

\newcommand{\calReD}{\mathcal{(R_{e},\Delta)}}
\newcommand{\ORe}{{R}_{e\lambda}}
\newcommand{\ODxy}{Ed(X,Y)}
\newcommand{\Iplx}{I^{\lambda}(U,X)}
\newcommand{\Iplz}{I^{\lambda}(U,Z)}
\newcommand{\Ipax}{I^{1}(U,X)}
\newcommand{\Ipaz}{I^{1}(U,Z)}
\newcommand{\Ipbx}{I^{2}(U,X)}
\newcommand{\Ipbz}{I^{2}(U,Z)}

\newcommand{\eql}{\stackrel{\triangle}{=}}
\newcommand{\PRD}{P^{*}(Y|X)}
\newcommand{\PRDs}{P^{*}(y|x)}
\newcommand{\Prd}{\widetilde{P}(Y|X)}
\newcommand{\Prds}{\widetilde{P}(y|x)}
\newcommand{\Ir}{I^{*}(Y|X)}
\newcommand{\Ird}{\widetilde{I}(Y|X)}
\newcommand{\Irrd}{I_{\lambda}(Y|X)}
\newcommand{\dap}{\stackrel{\cdot}{\approx}}
\newcommand{\epk}{\delta_{2}}
\newcommand{\Rek}{R_{e}^{*}}
\newcommand{\Rck}{R_{c}^{*}}
\newcommand{\calFA}{\mathcal{F}_{\bx}}
\newcommand{\pzsa}{P_{1}}

\newcommand{\PX}{P_{X}}
\newcommand{\PY}{P_{Y}}
\newcommand{\PW}{P_{W}}
\newcommand{\PXY}{P_{XY}}
\newcommand{\PXYZ}{P_{XYZ}}
\newcommand{\PVgU}{P_{V|U}}
\newcommand{\PBgA}{P_{B|A}}
\newcommand{\PYgX}{P_{Y|X}}

\newcommand{\Tga}{T_{P_{A}}^{\delta}}
\newcommand{\Tgx}{T_{P_{X}}^{\delta}}
\newcommand{\Tgu}{T_{P_{U}}^{\delta}}
\newcommand{\Tgy}{T_{P_{Y}}^{\delta'}}
\newcommand{\Tgba}{T_{P_{B|A}}^{\delta}(\ba)}
\newcommand{\Tgyx}{T_{P_{Y|X}}^{\delta}(\bx)}
\newcommand{\Tgux}{T_{P_{U|X}}^{\delta}(\bx)}
\newcommand{\Tguy}{T_{P_{Y|U}}^{\delta}(\bu)}
\newcommand{\Cu}{\mathcal{C(\bu)}}
\newcommand{\CU}{\mathcal{C}(\bU(v,k))}
\newcommand{\Cui}{\mathcal{C}(\bu(v,k))}

\newcommand{\Tgwa}{T_{P_{W_{1}}}^{\delta}}
\newcommand{\Tgwbwa}{T_{P_{W_{2}|W_{1}}}^{\delta}(\bw_{1})}
\newcommand{\Tgxwa}{T_{P_{XW_{1}}}^{2\delta}}
\newcommand{\Tgxwav}{T_{P_{XW_{1}V}}^{3\delta}}
\newcommand{\Tgxwavwb}{T_{P_{XW_{1}VW_{2}}}^{3\delta}}
\newcommand{\Tgxwavwc}{T_{P_{XW_{1}VW_{3}}}^{3\delta}}
\newcommand{\Tgxwavwbwc}{T_{P_{XW_{1}VW_{2}W_{3}}}^{4\delta}}
\newcommand{\Tgxwavwbwcwd}{T_{P_{XW_{1}VW_{2}W_{3}W_{4}}}^{5\delta}}

\newcommand{\Tgxzhx}{T_{P_{XZ\hat{X}}}^{5\delta|\calW_{1}\times\calW_{2}|}}
\newcommand{\Tgxytx}{T_{P_{XY\tilde{X}}}^{5\delta|\calW_{1}\times\calW_{2}|}}

\newcommand{\Tgxy}{T_{P_{XY}}^{\delta}}
\newcommand{\Tgxu}{T_{P_{XU}}^{\delta}}
\newcommand{\Tgxuy}{T_{P_{XUY}}^{\delta}}
\newcommand{\Tgxuyz}{T_{P_{XUYZ}}^{\delta}}
\newcommand{\Tgyz}{T_{P_{YZ}}^{\delta}}
\newcommand{\Tguz}{T_{P_{UZ}}^{\delta}}
\newcommand{\Tgxyt}{T_{P_{XY}}^{\tilde{\delta}}}

\newcommand{\calF}{\mathcal{F}}
\newcommand{\calBa}{\calB_{m}^{1,2}}
\newcommand{\calBb}{\calB_{m}^{3}}
\newcommand{\calQ}{\mathcal{Q}}
\newcommand{\calW}{\mathcal{W}}

\newcommand{\UT}{\widetilde{U}}
\newcommand{\uT}{\widetilde{u}}

\newcommand{\UTT}{\widetilde{U}'}
\newcommand{\calP}{\mathcal{P}}
\newcommand{\Q}{Q}

\begin{abstract}
Consider a source that produces independent copies of a triplet of
jointly distributed random variables,
$\{X_{i},Y_{i},Z_{i}\}_{i=1}^{\infty}$. The process $\{X_{i}\}$ is
observed at the encoder, and is supposed to be reproduced at two
decoders, decoder Y and decoder Z , where $\{Y_{i}\}$ and
$\{Z_{i}\}$ are observed, respectively, in either a causal or
non-causal manner. The communication between the encoder and the
decoders is carried in two successive stages. In the first stage,
the transmission is available to both decoders and they reconstruct
the source according to the received bit-stream and the individual
side information ($\{Z_{i}\}$ or $\{Y_{i}\}$). In the second stage,
additional information is sent to both decoders and they refine the
reconstructions of the source according to the available side
information and the transmissions at both stages. It is desired to
find the necessary and sufficient conditions on the communication
rates between the encoder and decoders, so that the distortions
incurred (at each stage) will not exceed given thresholds. For the
case of non-degraded causal side information at the decoders, an
exact single-letter characterization of the achievable region is
derived for the case of pure source-coding. Then, for the case of
communication between the encoder and decoders carried over
independent memoryless discrete channels with random states known
causally/non-causally at the encoder and with causal side
information about the source at the decoders, a single-letter
characterization of all achievable distortion in both stages is
provided and it is shown that the separation theorem holds. Finally,
for non-causal degraded side information, inner and outer bounds to
the achievable rate-distortion region are derived. These bounds are
shown to be tight for certain cases of reconstruction requirements
at the decoders, thereby shading some light on the problem of
successive refinement with non-degraded side information at the
decoders.

\end{abstract}

\vspace{0.5cm}

\section {Introduction}
\label{Introduction}

We consider an instance of the multiple description problem, which
is successive refinement (SR) of information. The term ``successive
refinement of information" is applicable to systems where the
reconstruction of the source is done in a number of stages. In such
systems, a source is encoded by a single encoder which communicates
with either a single decoder or a number of decoders in a successive
manner. At each stage, the encoder sends some amount of information
about the source to the decoder of that stage, which also has access
to all previous transmissions. The decoder bases its reconstruction
on all available transmissions, and, possibly, on some additional
side information (SI). The quality of reconstruction at each stage
(at each decoder) is measured with respect to a predefined
distortion measure. In the case of pure source coding, the
information transmitted by the encoder at each stage arrives at the
decoder noiselessly, while in the case of noisy channels connecting
the encoder and decoders, the transmission received at the decoder
is corrupted and thus, joint source-channel coding should be
applied.

A number of works have dealt with the problem of successive
refinement \cite{Koshelev}-\cite{SMer04}, and the related problem of
hierarchical coding \cite{SMer03}-\cite{MaorMerhav06}. In
\cite{SMer04}, the problem of successive source coding was studied
for the Wyner-Ziv setting, i.e., when SI is available to each
decoder non-causally \cite{Wyner_Ziv}. The encoder transmits a
source sequence, $\bX$, to two decoders in two successive stages.
Necessary and sufficient conditions were provided in \cite{SMer04},
in terms of single-letter formulas, for the achievability of
information per-stage rates corresponding to given distortion levels
of each communication step. For the case of identical SI available
at all decoders, the two-stage coding scheme was extended to include
any finite number of stages. Also, conditions for a source to be
successively refinable with degraded SI were introduced in
\cite{SMer04} for the two-stage case. Generally speaking, the notion
of degraded SI means that the quality of SI available at the
decoders of later stages is better than that of earlier stages.

In \cite{TianDiggavi06}, the problem of successive refinement with
SI available non-causally at each decoder was studied from a
different viewpoint. Instead of considering per-stage communication
rates, the analysis of successive refinement was performed with
respect to cumulative (sum-) rates achievable at each stage, under
per-stage source restoration assumptions.  A single-letter
characterization of the achievable region with successive coding
sum-rates and distortions was provided for the case of degraded SI
at the decoders. It turned out that when the rate-sums are analyzed,
it is possible to characterize an achievable rate-distortion region
for any number of stages as long as the SI at the decoders is
degraded.

In \cite{MaorMerhav06}, the problem of successive refinement was
investigated for the case of SI available causally at the decoders.
It turned out that, unlike the above described non-causal settings,
when SI is available causally, the characterization of the
achievable per-stage rate-distortion region is possible without
constraining SI to be degraded.

The works reported in the field of successive refinement thus far
have considered refinement of information when the transmission at
each stage has been addressed to a single decoder. There are,
however, many applications where a single encoder conveys
information to several decoders in a single transmission. Heegard
and Berger \cite{HeegardBerger85} and Kaspi \cite{Kaspi} studied
independently the following scenario: a single encoder communicates
via a single transmission with two decoders one of which accesses
the transmission only, while the other has a non-causal access to
some SI correlated with the source. The source sequence should be
reconstructed at both decoders with a certain accuracy and, under
these distortion constraints, it is desired to reduce the
communication rate as much as possible.

The minimum achievable communication rate, i.e., the rate-distortion
function obtained for this setup is referred to as the
\emph{Heegard-Berger rate-distortion function}. It was also extended
in \cite{HeegardBerger85} to include a coding theorem for more that
two decoders, each having access to a different SI with a degraded
structure. Now, assume that there is a demand for a better
reconstruction at either one or both decoders, i.e., the source is
required to perform a multi-level successive refinement, still
communicating with all decoders via a single transmission. A
question of obvious interest is the following: is it possible to
characterize the achievable rate-distortion region for this
generalized problem of successive refinement?

In this work, we jointly extend the works of \cite{HeegardBerger85},
\cite{SMer04}, \cite{TianDiggavi06} and \cite{MaorMerhav06}.
Specifically, we study the scenario of two-decoders, two-stage
successive refinement of information, with SI available at all
decoders in either a causal\footnote{There are few reasons for our
interest in the scenario of causal SI at the decoders. The first
motivation is an attempt to include the concept of SR in zero-delay
sequential coding systems. Schemes with causal SI can be also viewed
as denoising systems, where each decoder performs SI sequential
filtering with the aid of rate-constrained information provided by
the encoder. Introducing SR to such systems is of practical
importance, as it simplifies the decoding process in the sense of
performing denoising of the SI symbols causally, in a number of
steps, rather than using the entire SI sequence.} or non-causal
manner. For the causal case, we provide a single-letter
characterization of the achievable rate-distortion region, which is
straightforwardly extendable to any number of decoders accessible in
each stage and any finite number of stages. For the case of
non-causal SI, we provide inner and outer bounds to the achievable
rate-distortion region for the case of degraded SI. Note that
although the SI is degraded at each stage, when both stages are
viewed jointly, SI is no longer degraded (same SI is used at both
stages and thus it is not longer possible to say that at the later
stage the SI is of better quality), and therefore this setting is of
particular interest. When considering the case of causal SI, we
provide the exact achievable region in terms of the per-stage rates,
while for the case of non-causal SI, we refer to the sum-rates. The
difficulty in characterizing the per-stage rates for a general
scheme here is similar to that faced in \cite{SMer04}.

For the case of causal SI we then extend the noise-free setting into
a problem of communication over noisy discrete memoryless channels
with random states known causally or non-causally at the encoder at
all stages of communication. We obtain a single-letter
characterization of the region of all achievable distortions for
both decoders at both stages of communication. This characterization
reveals that the separation principle is applicable for this
problem, i.e., it is possible to separately encode the source
sequence with a good SR source code and then to transmit the
obtained bitstreams with a good channel code at each stage of
communication, without losing asymptotic optimality. This part of
the paper extends the results of \cite{ShamaiVerduZamir} and
\cite{MerShamai} to the multi-stage multi-decoder communication.
Specifically, in \cite{ShamaiVerduZamir} it was shown that the
separation principle holds for a single-stage single encoder-decoder
communication over a simple discrete memoryless channel. This
setting has been extended in \cite{MerShamai} to communication over
a channel with random parameters known causally or non-causally at
the encoder and decoder having non-causal access to the SI
correlated with the source and there also it was shown that separate
source channel coding is, in fact, optimal.

Note that all known closed form (single-letter) results regarding SR
(and its variations) for decoders having non-causal access to
different SI data, such as \cite{HeegardBerger85}, \cite{Rimoldi},
\cite{SMer04} and \cite{TianDiggavi06}, treat the case of degraded
SI at the decoders. Thus, there is a special interest in the
following sub-case of the problem treated in this paper - SR with
non-causal degrades SI at the decoders, when decoders are accessed
in the reversed order of degradedness of SI. Specifically, for the
two-stage scheme, assume that in the first stage some information is
to be conveyed to the decoder that has access to SI of a better
quality. Then, at the refinement stage, the decoder with less
informative SI should reconstruct the source sequence based on the
transmissions of both stages. This problem has been also addressed
in \cite{TianDiggavi07}. Specifically, in \cite{TianDiggavi07},
inner and outer bounds on the achievable rates and distortions have
been derived and it was shown that these bounds coincide when
reconstruction at either stage should be lossless at the matching
decoder. The work presented in \cite{TianDiggavi07} has been
performed in parallel to the researched described in this paper and
the inner bounds presented in \cite{TianDiggavi07} can be easily
derived from the results of this paper. The outer bound provided in
this paper is more precise than that provided in
\cite{TianDiggavi07} as is discussed in detail in Section
\ref{Main_Result}.

The outline of the paper is as follows: In Section \ref{notation},
we give notation conventions used throughout the paper. A formal
definition of the problem is provided in Section
\ref{System_Description}. In Section \ref{Main_Result}, for the case
of causal SI at the decoders, we give the exact characterizations of
the achievable rate-distortion region and formulate the coding
theorems for the successive-refinement two-stage source coding and
the joint source-channel coding;  for the case of non-causal SI at
the decoders, we provide inner and outer bounds to the
rate-distortion region and show that in some cases these bounds are
tight. The proofs are provided in Sections \ref{Proofs_C} and
\ref{Proofs_NC} for the cases of causal and non-causal SI,
respectively.

%-------------------------------------------------------------------
\section {Notation Conventions and Preliminaries}
\label{notation}
%-------------------------------------------------------------------

Throughout the paper, random variables will be denoted by capital
letters, specific values they may take will be denoted by the
corresponding lower case letters, and their alphabets will be
denoted by calligraphic letters. Similarly, random vectors, their
realizations, and their alphabets will be denoted, respectively, by
boldface capital letters, the corresponding boldface lower case
letters, and calligraphic letters, superscripted by the dimensions.
The notations $x_{i}^{j}$ and $X_{i}^{j}$, where $i$ and $j$ are
integers and $i\leq j$, will designate segments $(x_{i},...,x_{j})$
and $(X_{i},...,X_{j})$, respectively, where for $i=1$, the the
subscript will be omitted. For example, a random vector
$\bX=X_{1}^{N}=(X_{1},...,X_{N})$, ($N$-positive integer) may take a
specific vector value $\bx=x_{1}^{N}=(x_{1},...,x_{N})$ in
$\calX^{N}$, the $N$th order Cartesian power of $\calX$, which is
the alphabet of each component of this vector. The cardinality of a
finite set $\calA$ will be denoted by $|\mathcal{A}|$.

Sources and channels will be denoted generically by the letter $P$,
subscripted by the name of the random variable and its conditioning,
if applicable, e.g., $P_{X}(x)$ is the probability of $X=x$,
$P_{Y|X}(y|x)$ is the conditional probability of $Y=y$ given $X=x$,
and so on. Whenever clear from the context, these subscripts will be
omitted. The class of all discrete memoryless sources (DMSs) with a
finite alphabet $\mathcal{X}$ will be denoted by $\mathcal{P(X)}$,
with $\PX$ denoting a particular DMS in $\mathcal{P(X)}$, i.e., $
\mathcal{P(X)} = \{\PX:
    \sum_{x \in \calX}\PX(x)=1; \textrm{ }\textrm{ } \forall x\in
    \calX: \textrm{ }\textrm{ } \PX(x) \geq 0\}.$
For a given positive integer $N$, the probability of an $N$-vector
$\bx=(x_{1},...,x_{N})$ drawn from a DMS $P_{X}$, is given by
\begin {equation}
\label{QProb}
   \Pr\{X_{i}=x_{i},\textrm{ } i=1,...,N\} = \prod_{i=1}^{N}\PX(x_{i})
    \stackrel{\triangle}{=} \PX(\bx).
\end {equation}

A Markov chain formed by a triplet of random variables (RVs)
$(X,Y,Z)$ with a joint distribution $P_{XYZ}(x,y,z)$ will be denoted
by $X \div Y \div Z$.

A distortion measure (or distortion function) is a mapping from the
set $\mathcal{X}\times\mathcal{Y}$ into the set of non-negative
reals: $d: \mathcal{X}\times\mathcal{Y} \rightarrow
\mathcal{R}^{+}$. The additive distortion $d(\bx,\by)$ between two
vectors $\bx \in \calX^{N}$ and $\by \in \calY^{N}$ is given by: $
d(\bx,\by) = \frac{1}{N}\sum_{i=1}^{N}d(x_{i},y_{i}).$

The information-theoretic quantities, used throughout this paper,
are denoted using the conventional notations \cite{CT91}: For a pair
of discrete random variables $(X,Y)$ with a joint distribution
$P_{XY}(x,y)=P_{X}(x)P_{Y|X}(y|x)$, the entropy of $X$ is denoted by
$H(X)$, the joint entropy - by $H(X,Y)$, the conditional entropy of
$Y$ given $X$ - by $H(Y|X)$, and the mutual information by $I(X;Y)$,
etc., where logarithms are defined to the base 2.

We next describe the notation related to the method of types, which
is used throughout this paper in the direct proofs. For a given
memoryless source $\PX$ and a vector $\bx \in \calX^{N}$, the
empirical probability mass function is a vector
$P_{\mathbf{x}}=\{P_{\bx}(a),a \in \calX\}$, where
$P_{\mathbf{x}}(a)$ is the relative frequency of the letter $a \in
\calX$ in the vector $\bx$. For a scalar $\delta > 0$, the set
$\Tgx$ of all $\delta$-typical sequences is the set of the sequences
$\bx \in \calX^{N}$ such that $
\left|P_{\mathbf{x}}(a)-\PX(a)\right|\leq\delta $ for every $a \in
\calX$. In this paper, we use some known results from \cite{CT91}.
First, for every $\bx \in \Tgx$,
\begin {equation}
\label{TgxSize}
    2^{-N[H(X)+\epsilon_1]} \leq
    P_{X}(\bx) \leq 2^{-N[H(X)-\epsilon_1]},
\end{equation}
where $\epsilon_1 = \epsilon_{1}(\delta)$ vanishes as $\delta
\rightarrow 0$ and $N \rightarrow \infty$. It is also well-known (by
the weak law of large numbers) that:
\begin {equation}
\label{PrTgx}
    \Pr \big\{ \bX \notin \Tgx \big\} \leq \epsilon_2
\end {equation}
where $\epsilon_2 = \epsilon_{2}(\delta)$, $\epsilon_2\rightarrow 0$
as $N\to\infty$.

For a given conditional distribution $\PYgX$ and for each $\bx \in
\Tgx$, the set $\Tgxyt$ of all sequences $\by$ that are jointly
$\delta$-typical with $\bx$, is the set of all $\by$ such that:
\begin{eqnarray}
\left|P_{\mathbf{xy}}(a,b)-P_{\mathbf{x}}(a)\PYgX (b|a)\right|\leq
\tilde{\delta}
\end{eqnarray}
for all $a \in \calX, b \in \calY$, where $P_{ \mathbf{xy}}(a,b)$
denotes the fraction of occurrences of the pair $(a,b)$ in
$(\bx,\by)$. For any $\bx\in\Tgx$ and any $\tilde{\delta}>\delta$,
\begin {equation}
\label{TgyxSize}
    2^{-N[I(X;Y)+\epsilon_{3})]} \leq
    \sum_{\bold{y}:(\bold{x},\bold{y})\in\Tgxyt}P_{Y}(\bold{y}) \leq 2^{-N[I(X;Y)-\epsilon_{3}]},
\end{equation}
where $\epsilon_{3}=\epsilon_{3}(\delta,\tilde{\delta})$ vanishes as
$\delta,\tilde{\delta}\to 0$ and $N\to\infty$. These typicality
definitions and properties, are straightforwardly extendable for
jointly typical sequences which come in triplets, quadruplets and so
on and we use these in the paper.

%-------------------------------------------------------------------
\section {System Description and Problem Definition}
\label{System_Description}
\begin{figure}[!htbp]
 \begin{center}
  \resizebox{0.8\textwidth}{!}
     {\includegraphics{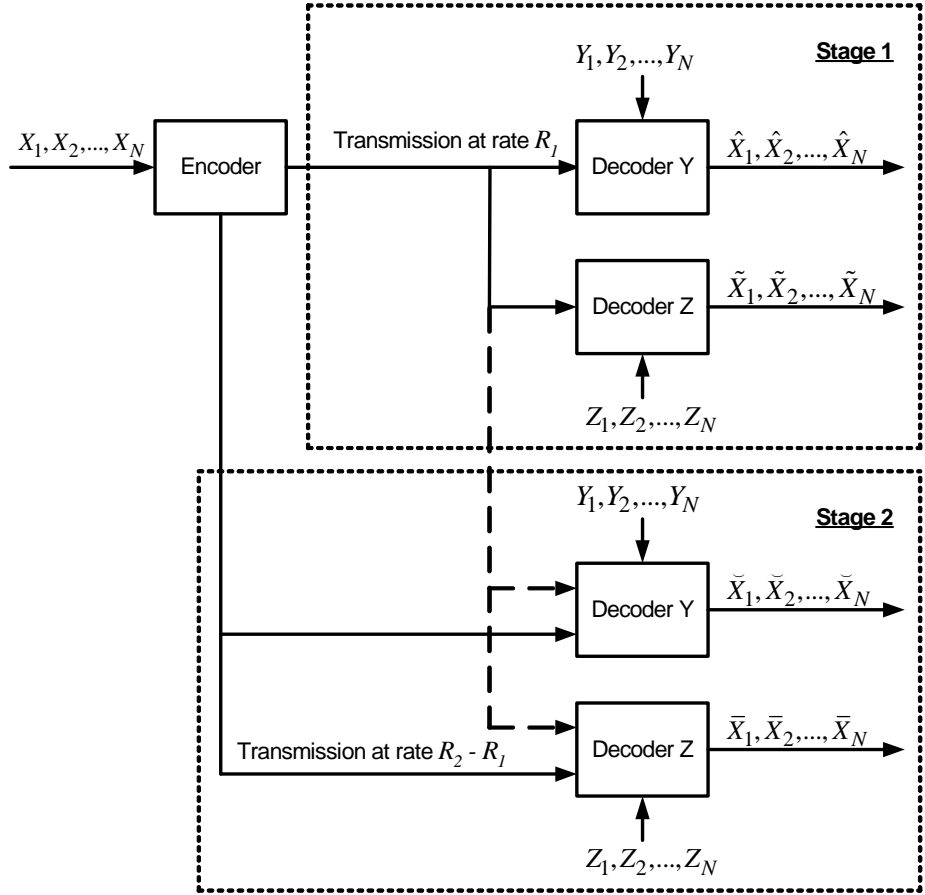}}
 \end{center}\vspace{-10pt}
\caption{Two-stage communication scheme.} \label{general}
\vspace{-3pt}
\end{figure}

%-------------------------------------------------------------------
\noindent We refer to the communication system depicted in Figure
\ref{general}.
 Consider a source that produces independent copies $\{X_{i},Y_{i},Z_{i}\}_{i\geq 1}$
of a triplet of RV's, $(X,Y,Z)$, taking values in a finite alphabet
$\calX\times\calY\times\calZ$, and drawn under a joint distribution
$\PXYZ$. The process $\{X_{i}\}$ is observed at the encoder and is
supposed to be reproduced at the decoders, where $\{Y_{i}\}$ and
$\{Z_{i}\}$ are observed at decoders Y and Z, respectively. The
source is available at the encoder non-causally and at the decoders
either causally or non-causally, at all stages. At the first stage of SR,
the reproductions at decoders Y and Z take values in the finite sets,
$\hat{\calX}$ and $\tilde{\calX}$, respectively, while at the second stage, the
reproduction finite sets are $\check{\calX}$ and $\bar{\calX}$, respectively.
%\footnote{The reconstructions of the
%first and the second stage at each of the decoders do not have to
%take values in the same sets, but this does not affect
%significantly the generality of the results but merely reduces the
%number of notations used in the paper. The extension
%to the case of different alphabets is straightforward.}.
%The source operates at the rate of $\rho_{s}$ symbol triples per second, i.e.,
%if it takes $T$ seconds to generated a block of $N$ source symbols, where $N$ is a positive integer, $N=\rho_s T$.

The coding scheme with causal/non-causal SI at the decoders operates
as follows: at the first transmission, the encoder sends some amount
of information to both decoders over the channel. We consider block
coding, i.e., an $N$-vector $\bX$ ($N$ is a positive integer) is
encoded at rate $R_{1}$ into a binary sequence
of length $M_1$, where $R_{1} = \frac{1}{N}\log_{2}{M_1}$. The binary sequence then takes values in $\{0,1,...,2^{NR_{1}}-1\}$. % Let $N = \rho_s T$ be a positive
%integer, where $T$ is the duration of the block in seconds.
At the first stage, when non-causal SI is considered, decoder Y
receives the binary bitstream and reconstructs
$\hat{\bX}=(\hat{X}_{1},...,\hat{X}_{N}) \in\hat{\calX}^{N}$, based
on it and the SI $\bY$, while in the case of causal SI, the
reconstruction of the $i$-th component, $\hat{X}_{i}$, is based on
the encoder transmission and only $i$ first symbols of the SI, i.e.,
$Y_{1}^{i}$. Similarly, with non-causal SI, decoder Z uses the
encoder transmission and $\bZ$ in its entirety and reproduces
$\tilde{\bX}=(\tilde{X}_{1},...,\tilde{X}_{N})
\in\tilde{\calX}^{N}$, while in the case of causal SI, only the
bitstream and $Z_{1}^{i}$ are used for reproduction of
$\tilde{X}_{i}$. The quality of reconstruction at each of the
decoders is judged in terms of the expectations of additive
distortion measures $
d_{y,1}(\bX,\hat{\bX})=\frac{1}{N}\sum_{i=1}^{N}d_{y,1}(X_{i},\hat{X}_{i})
$ and $
d_{z,1}(\bX,\tilde{\bX})=\frac{1}{N}\sum_{i=1}^{N}d_{z,1}(X_{i},\tilde{X}_{i}),
$ where $d_{y,1}(X,\hat{X})$ and $d_{z,1}(X,\tilde{X})$,
$X\in\calX$, $\hat{X}\in\hat{\calX}$, $\tilde{X}\in\tilde{\calX}$,
are non-negative, bounded distortion measures. At the second stage,
the encoder sends, at rate $R_{2}-R_{1}$, an additional information
about the source sequence to both decoders, also in the form of a
binary bitstream, this time of length $M_2 \eql 2^{N(R_2 - R_1)}$,
taking values in $\{0,1,...,2^{N(R_{2}-R_{1})}-1\}$ . The decoders
reconstruct the source sequence with better accuracy (in terms of
the distortion measures) according to both transmissions of the
encoder and the individual SI's. The distortions measures used at
the decoders Y and Z at this stage are also additive,
$d_{y,2}(\bX,\check{\bX})=\frac{1}{N}\sum_{i=1}^{N}d_{y,2}(X_{i},\check{X}_{i})
$ and $
d_{z,2}(\bX,\bar{\bX})=\frac{1}{N}\sum_{i=1}^{N}d_{z,2}(X_{i},\bar{X}_{i}),
$ where $d_{y,2}(X,\check{X})$ and $d_{z,2}(X,\bar{X})$,
$X\in\calX$, $\check{X}\in\check{\calX}$, $\bar{X}\in\bar{\calX}$,
are non-negative, bounded distortion measures. This setting can be
straightforwardly extended to any number of refinement stages as
well as any number of decoders at each stage. We confine ourselves
to the case of two decoders and two stages.
We begin with the case of non-causal SI. %observed by the decoders and
%formally define the following:
\begin{Definition}
\label{def3} An
$(N,M_{1},M_{2},\{\Delta_{y,k}$,$\Delta_{z,k}\}_{k=1}^{2})$ source
code for a single encoder, two decoders and two-stage successive
refinement with non-causal SI at the decoders, for the source
$\PXYZ$, consists of a first-stage encoder-decoder triplet
$(f_{1},g_{y,1},g_{z,1})$:\vspace{-8pt}
\begin{eqnarray}
    &&f_{1}: \calX^{N} \rightarrow \{1,2,...,M_{1}\},\\
    &&g_{y,1}: \calY^{N} \times \{1,2,...,M_{1}\} \rightarrow
    \hat{\calX}^{N},\\
    &&g_{z,1}: \calZ^{N} \times \{1,2,...,M_{1}\} \rightarrow \tilde{\calX}^{N},
\end{eqnarray}
and a second-stage encoder-decoder triplet
$(f_{2},g_{y,2},g_{z,2})$:
\begin{eqnarray}
    &&f_{2}: \calX^{N} \rightarrow \{1,2,...,M_{2}\},\\
    &&g_{y,2}: \calY^{N} \times \{1,2,...,M_{1}\} \times \{1,2,...,M_{2}\} \rightarrow
    \check{\calX}^{N},\\
    &&g_{z,2}: \calZ^{N} \times \{1,2,...,M_{1}\} \times \{1,2,...,M_{2}\} \rightarrow \bar{\calX}^{N},
\end{eqnarray}
such that
\begin{eqnarray}
Ed_{y,1}(\bX,\hat{\bX}) \leq N\Delta_{y,1} \textrm{ }\textrm{
}\textrm{ }\textrm{ }\textrm{ }\textrm{ } Ed_{z,1}(\bX,\tilde{\bX})
\leq N\Delta_{z,1}\nonumber
\end{eqnarray}
and
\begin{eqnarray}
Ed_{y,2}(\bX,\check{\bX}) \leq N\Delta_{y,2} \textrm{
}\textrm{ }\textrm{ }\textrm{ }\textrm{ }\textrm{ }
Ed_{z,2}(\bX,\bar{\bX}) \leq N\Delta_{z,2}.\nonumber
\end{eqnarray}
%$Ed_{y,1}(\bX,\hat{\bX}) \leq N\Delta_{y,1}$,
%$Ed_{z,1}(\bX,\tilde{\bX}) \leq N\Delta_{z,1}$,
%$Ed_{y,2}(\bX,\check{\bX}) \leq N\Delta_{y,2}$ \\and
%$Ed_{z,2}(\bX,\bar{\bX}) \leq N\Delta_{z,2}$.
%\vspace{-8pt}
%\begin{eqnarray}
%%\label{ineq11}
%Ed_{1}(\bX,g_{1,i}) \leq N\Delta_{1,i}, \textrm{ }\textrm{ and
%}\textrm{ }
%%\label{ineq22}
%Ed_{2}(\bX,g_{2,i}) \leq N\Delta_{2,i}.
%\end{eqnarray}
\end{Definition}
When SI is available to the decoders causally, in analogy to
Definition \ref{def3}, it is possible to define an
$(N,M_{1},M_{2},\{\Delta_{y,k}$,$\Delta_{z,k}\}_{k=1}^{2})$, source
code for coding with causal SI, where the first-stage decoder pair
$(g_{y,1},g_{z,1})$ is now presented via $\{g_{y,1,i}\}_{i=1}^N$ and
$\{g_{z,1,i}\}_{i=1}^N$, where $g_{y,1,i}$ and $g_{z,1,i}$ denote
the reconstruction functions for the $i-th$ symbol of $\hat{X}^{N}$
and $\tilde{X}^{N}$, respectively:
\begin{eqnarray}
    &&g_{y,1,i}: \calY_1^{i} \times \{1,2,...,M_{1}\} \rightarrow
    \hat{\calX},\\
    &&g_{z,1,i}: \calZ_1^{i} \times \{1,2,...,M_{1}\} \rightarrow
    \tilde{\calX}.
\end{eqnarray}
Similar adjustments of definitions should be applied to the second
stage, considering now $(g_{y,2},g_{z,2})$ presented in terms of
$\{g_{y,2,i}\}_{i=1}^N$ and $\{g_{z,2,i}\}_{i=1}^N$:
\begin{eqnarray}
    &&g_{y,2,i}: \calY_1^{i} \times \{1,2,...,M_{1}\}\times \{1,2,...,M_{2}\} \rightarrow
    \check{\calX},\\
    &&g_{z,2,i}: \calZ_1^{i} \times \{1,2,...,M_{1}\}\times \{1,2,...,M_{2}\} \rightarrow
    \bar{\calX}.
\end{eqnarray}
The sum-rate pair $(R_{1},R_{2})$ of the
$(N,M_{1},M_{2},\{\Delta_{y,k}$,$\Delta_{z,k}\}_{k=1}^{2})$ code for
two stage successive refinement for two decoders is given by
$R_{1}=\frac{1}{N}\log_{2}(M_{1})$ and
$R_{2}=\frac{1}{N}\log_{2}(M_{1}\cdot M_{2})$.

\begin{Definition}
\label{def4} Given a distortion quadruplet $\bD =
\{\Delta_{y,k},\Delta_{z,k}\}_{k=1}^{2}$, a rate pair
$(R_{1},R_{2})$ is said to be achievable with SI $(Y,Z)$ if for
every $\epsilon > 0$, there exists a sufficiently large block length
$N$, for which there is an $(N,2^{N(R_{1}+\epsilon)}$,
$2^{N(R_{2}+\epsilon)}$, $\Delta_{y,1}+\epsilon$,
$\Delta_{z,1}+\epsilon$ ,$\Delta_{y,2}+\epsilon$,
$\Delta_{z,2}+\epsilon)$, source code for successive refinement with
non-causal SI at the decoders for the source $\PXYZ$.
\end{Definition}

The definition of the notion of an achievable region with causal SI
\emph{per-stage} rates can be straightforwardly modified in parallel
to Definition \ref{def4}, referring to the first stage rate $R_{1}$
and the second-stage rate $ \Delta R = R_{2}-R_{1} =
\frac{1}{N}\log_{2}(M_{2})$. The collection of all $\bD$-achievable
rate pairs is the achievable rate-region for successive-refinement
coding with non-causal (respectively, causal) SI and is denoted by
$\calR(\bD)_{nc}$ (respectively, $\calR(\bD)_{c}$). The collection
of all
$(R_{1},R_{2},\{\Delta_{1,k},\Delta_{2,k}\}_{k=1}^{2})$-achievable
rate-distortion tuples is the achievable rate-distortion region, and
is denoted by $\calR\calD_{nc}$ and $\calR\calD_{c}$, referring to
non-causal and causal settings, respectively. In this work, we
propose strategies for (asymptotically) achieving any given point in
$\calR\calD_{c}$ and certain points in $\calR\calD_{nc}$.

It is also interesting to investigate the scenario where
communication between the encoder and the decoders is carried over a
noisy media. In this case, the source block $\bX$ is fed into a
\emph{joint source-channel} encoder, whereas the corresponding
blocks of $\bY$ and $\bZ$ are fed as side information in either a
causal or non-causal manner into the Y and Z decoders, respectively.
In the sequel, we confine ourself to the case of causal source SI at
both decoders.\footnote{Since the complete characterization of
$\calR\calD_{nc}$ is still open, there is no point in analyzing the
scenario of communication over noisy channels for the case of
non-causal \emph{source} SI at the decoders.} In this paper, at each
stage of communication, the noisy media is modeled by a discrete
memoryless channel whose output is governed by its input and a
random parameter which is known at the encoder either causally or
non-causally.

\begin{figure}[!htbp]
 \begin{center}
  \resizebox{0.8\textwidth}{!}
     {\includegraphics{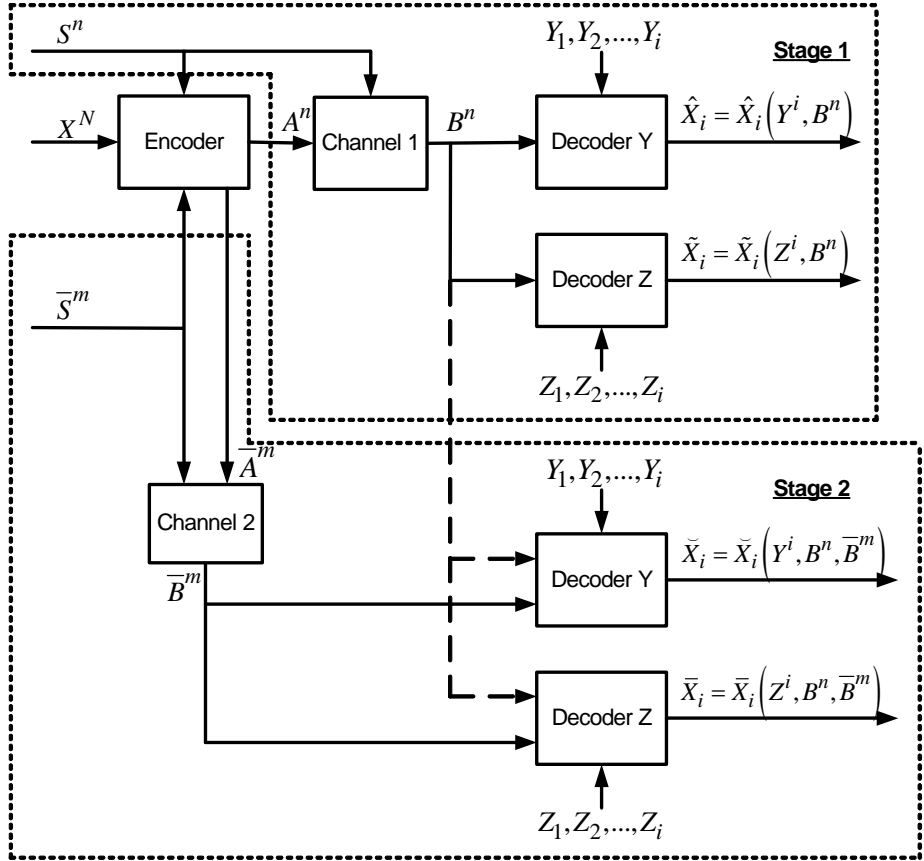}}
 \end{center}\vspace{-10pt}
\caption{Communication over noisy channels with causal SI.}
\label{general2} \vspace{-3pt}
\end{figure}

Consider the communication scheme depicted in Figure \ref{general2}.
The channel used at the first stage is channel 1, $P_{B|A,S}$, and
at the second stage is used channel 2,
$P_{\bar{B}|\bar{A},\bar{S}}$. The channels are independent and we
denote their capacities by $C_1$ and $C_2$, respectively. %At the
%duration of $T$ seconds, the channels 1 and 2 operate at rates
%$\rho_{c1}$ and $\rho_{c2}$ channel uses per second, respectively.
The channels work as follows: The input of Channel 1 is a vector
pair $(A^n_{1},S^n_{1})$, where $n$ is a positive integer and where
$A$ and $S$ take values in the finite sets, $\calA$ and $\calS$,
respectively. Channel 1 produces a vector output $B^n$, whose
components take values in the finite set $\calB$. The conditional
probability of $(B^n)$ given $(A^n,S^n)$ is characterized by
$P_{B^n|A^n,S^n}(b^n|a^n,s^n)=\prod_{i=1}^nP_{B|A,S}(b_i|a_i,s_i)$.
The vector $A^n$ is referred to as the channel input and $S^n$ is
referred to as the channel state sequence, governed by another
discrete memoryless process
$P_{S^n}(s^n)=\prod_{i=1}^{n}P_{s}(s_i)$, independently of
$(X^N,Y^N,Z^N)$. %It is also assumed that $(Y^N,Z^N) \div X^N \div
%B^n$ is a Markov chain.
The operation of Channel 2 is described in a similar fashion by the
triplet $(\bar{A}^m,\bar{B}^m,\bar{S}^m,)$ instead of
$(A^n,B^n,S^n)$ and corresponding marginal and conditional
probabilities. Note that in the context of Channel 2, all blocks are
of length $m$, where $m$ is a positive integer. We denote the
source-channel rate ratios by $\rho_{1}\eql\frac{n}{N}$ and
$\rho_{2}\eql\frac{m}{N}$.

Now, instead of the binary bitstream generated in the noise-free
case, the first-stage joint source-channel encoder implements a
deterministic function $a^n=f_1(x^N,s^n)$ and the second-stage joint
source-channel encoder implements another deterministic function
$\bar{a}^m=f_2(x^N,\bar{s}^m)$. If the channel states are available
at the encoder causally, each channel symbol $a_i$ depends only on
$x^N$, $a^{i-1}$ and $s^i$, and each $\bar{a}_i$ depends only on
$x^N$, $\bar{a}^{i-1}$ and $\bar{s}^i$. In the non-causal case, each
channel symbol $a_i$ depends on $x^N$, $a^{i-1}$ and $s^n$, and each
$\bar{a}_i$ depends $x^N$, $\bar{a}^{i-1}$ and $\bar{s}^m$. The
first-stage decoders Y and Z are defined now by deterministic
functions $g_{y,1}(y^N,a^n)$ and $g_{z,1}(z^N,a^n)$, respectively,
and the second stage decoders Y and Z are defined by deterministic
functions $g_{y,2}(y^N,b^n,\bar{b}^m)$ and
$g_{z,2}(z^N,b^n,\bar{b}^m)$, respectively. The channel states $\bS$
and $\bar{\bS}$ are independent and we interpret the independence of
the channels via the Markov relation
$(\bS,\bB) \div \bX \div (\bar{\bS},\bar{\bB})$.%\footnote{This
%definition is less general than
%$\bar{\bB}\div(\bX,\bS,\bar{\bS})\div\bar{\bB}$ which, in part, does
%not require independence of $\bS$ and $\bar{\bS}$ in addition to
%independence of $\bX$ and state information at both channels.}.

In parallel to Definitions \ref{def3} and \ref{def4}, we define the
following:
\begin{Definition}
\label{def5} For a given memoryless source $\PXYZ$ and two
memoryless channels with random states $P_{B|A,S}$ and
$P_{\bar{B}|\bar{A},\bar{S}}$ an
$(N,n,m,\Delta_{y,1},\Delta_{z,1},\Delta_{y,2},\Delta_{z,2})$ joint
source-channel code for successive refinement with causal state
information at the encoder and causal side information at the
decoders consists of a sequence of $n$ first-stage encoding
functions:
\begin{eqnarray}
    f_{1,i}: \calX^{N}\times \calS^i \rightarrow \calA_{i}, \textrm{ }\textrm{ }\textrm{ }i=1,...,n,
\end{eqnarray}
a sequence of $N$ first-stage decoding functions
\begin{eqnarray}
    g_{y,1,i}: \calY^{i} \times \calB^{n} \rightarrow
    \hat{\calX}, \textrm{ }\textrm{ }\textrm{ }i=1,...,N,
\end{eqnarray}
and
\begin{eqnarray}
    g_{z,1,i}: \calZ^{i} \times \calB^{n} \rightarrow
    \tilde{\calX}, \textrm{ }\textrm{ }\textrm{ }i=1,...,N,
\end{eqnarray}
a sequence of $m$ second-stage encoder functions
\begin{eqnarray}
    f_{2,i}: \calX^{N}\times \mathcal{\bar{S}}^i \rightarrow
    \mathcal{\bar{A}}_{i}, \textrm{ }\textrm{ }\textrm{ }i=1,...,m,
\end{eqnarray}
and a sequence of $N$ second-stage decoding functions:
\begin{eqnarray}
    g_{y,2,i}: \calY^{i} \times \calB^{n} \times \mathcal{\bar{B}}^m \rightarrow
    \check{\calX}, \textrm{ }\textrm{ }\textrm{ }i=1,...,N,
\end{eqnarray}
and
\begin{eqnarray}
    g_{z,2,i}: \calZ^{i} \times \calB^{n} \times \mathcal{\bar{B}}^m \rightarrow
    \bar{\calX}, \textrm{ }\textrm{ }\textrm{ }i=1,...,N,
\end{eqnarray}
such that
\begin{eqnarray}
Ed_{y,1}(\bX,\hat{\bX}) \leq N\Delta_{y,1} \textrm{ }\textrm{
}\textrm{ }\textrm{ } Ed_{z,1}(\bX,\tilde{\bX}) \leq N\Delta_{z,1}
\nonumber
\end{eqnarray}
and
\begin{eqnarray}
Ed_{y,2}(\bX,\check{\bX}) \leq N\Delta_{y,2} \textrm{ }\textrm{
}\textrm{ }\textrm{ } Ed_{z,2}(\bX,\bar{\bX}) \leq N\Delta_{z,2},
\nonumber
\end{eqnarray}
where the expectations are w.r.t. the source and the channels.
\end{Definition}

\begin{Definition}
\label{def6} Given the source-channel rate ratios $\rho_{1}$ and
$\rho_{2}$, a distortion quadruplet $\bD =
\{\Delta_{y,k},\Delta_{z,k}\}_{k=1}^{2}$ is said to be achievable if
for every $\epsilon > 0$, there exist sufficiently large $N$, $n$
and $m$, with $\rho_1 = n/N$ and $\rho_2 = m/N$, and there exists an
$(N,n,m,\Delta_{y,1}+\epsilon,\Delta_{z,1}+\epsilon,\Delta_{y,2}+\epsilon,\Delta_{z,2}+\epsilon)$
joint source-channel code for successive refinement with
causal/non-causal state information at the encoder and causal side
information at the decoders for the source $\PXYZ$ and the channels
$P_{B|A,S}$, $P_{\bar{B}|\bar{A},\bar{S}}$. The distortion region,
denoted $\calD$, is the closure of the set of all achievable
quadruplets $\bD$.
\end{Definition}
We provide a single-letter characterization of $\calD$ for the cases
of causal/non-causal channel state information availability at the
encoder. In particular, we show that any given point in $\calD$ can
be achieved by separate source coding for the source $\PXYZ$
(achieving $\calR\calD_c$) and capacity-achieving channel coding
(independently of the source).
%
%The proof of the separation theorem is intimately related to the
%characterization of the channel capacity for a single input -
%multiple output (SIMO) channel\footnote{To the best of our
%knowledge, such characterization is not explicitly stated for a
%general SIMO channel with random parameters, but is available only
%for degraded channels and a set of messages to be conveyed to the
%receivers.}. For the completeness, we provide next the coding
%theorem.
%\begin{Theorem}
%\label{theo:broadcast}
%    The capacity of the broadcast channel with state
%    information available causally at the transmitter is the
%    following:
%    \begin{eqnarray}
%        C = \max_{p_{u}p_{s}p_{x|su}}\min\{I(U;Y_{1}),I(U;Y_{2})\},
%    \end{eqnarray}
%    where the auxiliary RV $U$ is of finite cardinality upper-bounded
%    by $|\calX|\times|\calS|+2$.
%\end{Theorem}
%The proof of Theorem \ref{theo:broadcast} is outlined in the
%Appendix.

%-------------------------------------------------------------------
\section{Main Result}
\label{Main_Result}
%-------------------------------------------------------------------
\subsection{Causal Side Information}
\subsubsection{Pure Source Coding}
We begin with the case where availability of SI at the decoders is
restricted to be causal. Let a distortion quadruplet $\bD\eql
(\{\Delta_{y,k},\Delta_{z,k}\}_{k=1}^{2})$ be given. Define
$\calR^{*}(\bD)_{c}$ to be the set of all rate pairs $(R_{1},R_{2})$
for which there exist RVs $(W_{1},W_{2})$, taking values in finite
alphabets, $\calW_{1}$,$\calW_{2}$, respectively, s.t the following
holds simultaneously:

\noindent 1. The following Markov chain holds:
\begin{eqnarray}
\label{theo_1gau2}
    (W_{1},W_{2}) \div X \div (Y,Z).
\end{eqnarray}
\noindent 2. There exist deterministic decoding functions
$G_{y,1}:\textrm{ } \calY\times\calW_{1}\rightarrow\hat{\calX}$,
$G_{z,1}:\textrm{ }\calZ\times\calW_{1}\rightarrow\tilde{\calX}$,
and $G_{y,2}:\textrm{ }
\calY\times\calW_{1}\times\calW_{2}\rightarrow\check{\calX}$,
$G_{z,2}:\textrm{
}\calZ\times\calW_{1}\times\calW_{2}\rightarrow\bar{\calX}$, such
that
\begin {eqnarray}
\label{theo_1cau2}
    Ed_{y,1}(X,G_{y,1}(Y,W_{1})) \leq \Delta_{y,1}
\end{eqnarray}
\begin{eqnarray}
 \label{theo_1dau2}
    Ed_{z,1}(X,G_{z,1}(Z,W_{1})) \leq \Delta_{z,1}
\end {eqnarray}
\begin {eqnarray}
\label{theo_1cau3c}
    Ed_{y,2}(X,G_{y,2}(Y,W_{1},W_{2})) \leq \Delta_{y,2}
\end{eqnarray}
\begin{eqnarray}
 \label{theo_1dau4}
    Ed_{z,2}(X,G_{z,2}(Z,W_{1},W_{2})) \leq \Delta_{z,2}
\end {eqnarray}

\noindent 3. The alphabets $\calW_{1}$ and $\calW_{2}$ satisfy:
\begin{eqnarray}
\label{theo_1eau2}
    |\calW_{1}| \leq |\calX| + 5, & |\calW_{2}| \leq |\calX|\cdot|\calW_{1}| + 2
\end{eqnarray}

\noindent 4. The rates $R_{1}$ and $R_{2}$ satisfy
\begin {eqnarray}
    \label{theo_1acau}
    R_{1} \geq I(X;W_{1}) & R_{2}-R_{1} \geq I(X;W_{2}|W_{1}).
\end{eqnarray}

\noindent The main result of this subsection is the following:
%\vspace{0.5\baselineskip}

\begin{Theorem} \label{theo_source_causal} For
any DMS $\PXYZ$,
    \begin{eqnarray}
        \calR(\bD)_{c} = \calR^{*}(\bD)_{c}.
    \end{eqnarray}
\end{Theorem}

The proof of Theorem \ref{theo_source_causal} appears in Section
\ref{Proofs_C}. Note that when SI is available at the decoders
causally, there is no degradedness assumption on SI, which is very
different from the case of SR with non-causal SI even when a single
decoder is considered at each stage
\cite{SMer04}-\cite{TianDiggavi06}, as well as for the multi-group
SR discussed in the next section. %Similarly, no special internal
%Markov relations characterize the auxiliary RVs $(W_1,W_2)$
%\footnote{It is possible to reformulate the Markov condition
%(\ref{theo_1gau2}) into $W_{1} \div W_{2} \div X \div (Y,Z)$, but
%this condition would be a by-product of a choice of auxiliary RVs in
%the converse proof rather than a ``must" condition in the
%achievability proof, as it happens in the case of non-causal SI.}.

The relative simplicity of characterization of $\calR(\bD)_{c}$ is
better understood when studying the achievability scheme: The direct
part is based on the fact that the encoder transmits a concatenation
of indexes of the auxiliary codewords \footnote{The direct use of
indexes of the auxiliary codewords, similarly as is done for coding
without SI at the decoder, was first introduced in
\cite{GamalWeissman}, in the achievability proof of the
characterization of the rate-distortion function with causal SI at
the decoder.} instead of bin numbers transmitted in the non-causal
setting \cite{SMer04}-\cite{TianDiggavi06}. Hence, each decoder can
access all the auxiliary codewords directly and, unlike in the
non-causal setting, it does not use its SI to retrieve codewords,
but only for reconstruction. %In fact, in this achievability scheme,
%some of the decoders can achieve source reconstruction of a better
%quality than specified by the distortion constraint, since all the
%decoders use the same encoder transmission which is shaped to
%satisfy all the decoders' requirements simultaneously. Thus, in
%general, decoders can ``decide" to utilize the transmission
%entirely, achieving a better than required reconstruction of the
%source, or to utilize the transmission partially, reducing the
%reconstruction complexity.
Unlike in the case of coding with
non-causal SI at the decoders, the results obtained for the
two-decoder two-stage coding with causal SI are straightforwardly
extendable to any number of decoders and refinement stages and the
number of auxiliary RVs is determined solely by the number of
communication stages\footnote{While, as we show in the next section,
in the non-causal setting, at each stage, for each decoder, at least
one auxiliary codeword is added to the direct scheme.}.

%Note that when SIs  are available causally, only two auxiliary RVs
%are used, reducing the system complexity. This case is similar but
%not identical to the case of successive refinement for two groups of
%decoders, where in both groups the encoders and decoders have access
%to identical SI and the quality of the SI available to the second
%group is better than that available to the first group. In that
%case, the first group of decoders receives a transmission which is
%designed so that the hardest constraint will be satisfied, while the
%second group receives an additional transmission which together with
%the first one should maintain the hardest distortion constraint of
%the second group. In our case, the transmission of each stage is
%designed so that the hardest distortion constraints of decoders with
%different SI are satisfied at each stage, so, in fact, here also
%some decoders have access to transmission of a better quality than
%needed for their use only and they can decide if to utilize the
%transmission entirely, achieving a better than required
%reconstruction of the source data, or to utilize the transmission
%partially, reducing reconstruction complexity.

%--------------------------------------------------------
%--------------------------------------------------------
%----------------The Separation Theorem------------------
%--------------------------------------------------------
%--------------------------------------------------------
\subsubsection {Joint Source-Channel Coding}

\label{separation_proof}

We next address the problem of joint source channel coding, where at
each communication stage the encoder conveys its information to two
decoders over a noisy stationary memoryless channel governed by a
random state, which is known causally or non-causally to the
encoder. The general scheme is described in Fig. \ref{general2}. The
necessary and sufficient conditions for $(\Delta_{1},\Delta_{2})$ to
be the achievable distortion levels are summarized in the following
Theorem:
\begin{Theorem}
\label{theo_separation} Given a DMS $\PXYZ$, the distortion levels
$(\{\Delta_{y,k},\Delta_{z,k}\}_{k=1}^{2})$ are achievable for
successively refinable communication with causal SI at the decoders
over noisy stationary memoryless channels $P_{B,O|A,S}$ and
$P_{\bar{B},\bar{O}|\bar{A},\bar{S}}$ with channel states known at
the encoder either causally or non-causally if and only if there
exist auxiliary RVs $W_{1}$ and $W_{2}$, taking values in finite
alphabets $\calW_{1}$ and $\calW_{2}$, of cardinalities given by
(\ref{theo_1eau2}) and satisfying (\ref{theo_1gau2}), and
deterministic decoding functions $G_{y,1}$, $G_{z,1}$, $G_{y,2}$ and
$G_{z,2}$, satisfying (\ref{theo_1cau2}) - (\ref{theo_1dau4}),
respectively, such that
\begin {eqnarray}
\label{theo_2a}
   I(X;W_{1}) \leq  \rho_1C_1,
\end {eqnarray}
\begin {eqnarray}
\label{theo_2b}
    I(X;W_{2}|W_{1}) \leq \rho_{2}C_{2}.
\end {eqnarray}
\end{Theorem}

\indent There is an obvious similarity between the characterization
of $\calD_{c}$ and the characterization of the region of all
achievable distortion levels described in Theorem
\ref{theo_separation}, both for the cases of causal and non-causal
state information at the encoder. The only difference in
characterizations is the following: in the case of communication
over noisy channel the upper-bounds in (\ref{theo_2a}) and
(\ref{theo_2b}) are $\rho_{1}C_{1}$ and $\rho_{2}C_{2}$, while in
the noise-free case, these bounds are substituted by $R_1$ and $R_2
- R_1$, respectively. Therefore, a possible achievability scheme is
the one based on separate source and channel coding.

The direct proof of Theorem \ref{theo_separation} comes from a
concatenation of the asymptotically optimal source code designed for
multi-group successive refinement, which is independent of the
channels, and a reliable channel codes, independent of the source,
designed for each of the channels (with channel state informations
available to the encoder either causally or non-causally). The
channel codes should achieve (at least asymptotically) the capacity
of the relevant channels. Now, if such source and channel codes are
used and the distortion constraints are maintained by the source
code, as soon as $I(X;W_1) \leq \rho_1 C_1$ and $I(X;W_2|W_1) \leq
\rho_2 C_2$, it is always possible to select source and channel
rates $R_{s1}$ and $R_{c1}$ for the first stage and $R_{s2}-R_{s1}$
and $R_{c2}$ for the second stage such that $NI(X;W_1)\leq NR_{s1} =
nR_{c1}\leq nC_1$ and $NI(X;W_2|W_1)\leq N[R_{s2}-R_{s1}] =
mR_{c2}\leq mC_2$. Now, it is possible to compress the source
sequence into $R_{s1}$ bits per symbol for the first stage and into
$R_{s2}-R_{s1}$ bits per symbol for the refinement stage, such that
the distortions $\{(\Delta_{y,j},\Delta_{z,j})\}_{j=1}^{2}$ are
satisfied and then map the obtained bitstreams of length $NR_{s1}$
and $N[R_{s2}-R_{s1}]$ into channel codewords of length $nR_{c1}$
and $mR_{c2}$, respectively. Since $R_{c1} \leq C_1$ and $R_{c2}
\leq C_2$, from the standard coding theorem (\cite{GP80} or
\cite{Shannon}), there exist channel codes that cause asymptotically
negligible distortions. Also, by the source coding theorem (Theorem
\ref{theo_source_causal}) all the distortions for which
$NI(X;W_1)\leq NR_{s1}$ and $NI(X;W_2|W_1)\leq N[R_{s2}-R_{s1}]$ are
achievable. Thus, the distortions
$\{(\Delta_{y,j},\Delta_{z,j})\}_{j=1}^{2}$ such that $NI(X;W_1)\leq
nC_1$ and $NI(X;W_2|W_1)\leq mC_2$ are achievable.
%In such a scheme, the source code is designed independently of the
%channels for the two-stage two-decoders system with causal SI and
%the binary data-streams produces as an output of such source coding
%are then encoded with good channel codes, designed independently for
%each of the channels.
The details of the converse proof are provided in Section
\ref{Proofs_C}, and, similarly as in the noise-free case, the proof
is easily extendable to more than two communication stages
and more than two decoders at each stage.%\footnote{The statement is

\subsection {Non-Causal Degraded Side Information}
\label{Main_Result_Multiple_Description} Unlike in the case of
causal SI, in the noncausal case, a closed-form characterization of
the achievable rate-distortion region with non-causal SI at the
decoders is yet to be derived. In this subsection, we provide outer
and inner bounds to the achievable region, discuss the differences
between the bounds and show that in certain cases, the bounds
coincide, i.e., the rate-distortion region is fully characterized
for these special cases. We begin with the outer bound.

\subsubsection{Outer Bound}
\label{Outer_Bound_def} Define $\calR^{**}(\bD)_{nc}$ to be the set
of all rate pairs $(R_{1},R_{2})$ for which there exist RVs
$\{W_{i}\}_{i=1}^{4}$ and $V$, taking values in finite alphabets,
$\{\calW_{i}\}_{i=1}^{4}$ and $\calV$, respectively, such that
(s.t.) the following conditions are satisfied:

\noindent 1. \begin{eqnarray}\label{Markov_non-causal-outer-bound}
(W_{1},W_{2},W_{3},W_{4},V) \div X \div Z \div Y \end{eqnarray} is a
Markov chain.

\noindent 2. There exist deterministic decoding functions
$G_{y,1}:\textrm{ } \calY\times\calW_{1}\rightarrow\hat{\calX}$,
$G_{z,1}:\textrm{ }\calZ\times\calW_{1}\times\calW_{2}\times\calV
\rightarrow\tilde{\calX}$, $G_{y,2}:\textrm{ } \calY\times\calW_{1}
\times\calW_{3}\times\calV\rightarrow\check{\calX}$ and
$G_{z,2}:\textrm{
}\calZ\times\calW_{1}\times\calW_{2}\times\calW_{3}
\times\calW_{4}\times\calV \rightarrow\bar{\calX}$, such that
\vspace{-3pt}
\begin {eqnarray}
\label{theo_1c3ll}
   & Ed_{y,1}(X,G_{y,1}(Y,W_{1})) \leq \Delta_{y,1}\\
 \label{theo_1d3}
   & Ed_{z,1}(X,G_{z,1}(Z,W_{1},W_{2},V)) \leq \Delta_{z,1}\\
\label{theo_1c3}
   & Ed_{y,2}(X,G_{y,2}(Y,W_{1},W_{3},V)) \leq \Delta_{y,2}\\
 \label{theo_1d3a}
   & Ed_{z,2}(X,G_{z,2}(Z,W_{1},W_{2},W_{3},W_{4},V)) \leq \Delta_{z,2}
\end {eqnarray}

\noindent 3. The alphabets $\{\calW_{k}\}_{k=1}^{4}$ and $\calV$
satisfy:
\begin{eqnarray}
\label{theo_1e3}
    |\calW_{1}| \leq |\calX| + 5,
\end{eqnarray}
\begin{eqnarray}
\label{theo_1f3}
    |\calV| \leq |\calX|\cdot|\calW_{1}| + 4,
\end{eqnarray}
\begin{eqnarray}
\label{theo_1gg3}
    |\calW_{2}| \leq |\calX|\cdot|\calW_{1}|\cdot|\calV| + 3,
\end{eqnarray}
\begin{eqnarray}
\label{theo_1g3}
    |\calW_{3}| \leq
    |\calX|\cdot|\calW_{1}|\cdot|\calW_{2}|\cdot|\calV| + 2,
\end{eqnarray}
\begin{eqnarray}
\label{theo_1h3}
    |\calW_{4}| \leq
    |\calX|\cdot|\calW_{1}|\cdot|\calW_{2}|\cdot|\calW_{3}|\cdot|\calV|
    + 1.
\end{eqnarray}

\noindent 4. The rates $R_{1}$ and $R_{2}$ satisfy
\begin {eqnarray}
    \label{theo_1a3aa}
    R_{1} \geq I(X;W_{1}|Y)+I(X;W_{2},V|W_{1},Z)
\end{eqnarray}
\begin{eqnarray}
\label{theo_1b3aa}
    R_{2} &\geq& I(X;W_{1},W_{3},V|Y)+I(X;W_{2},W_{4}|W_{1},W_{3},V,Z)
\end {eqnarray}

\noindent The outer bound to the rate-distortion region is
summarized in the following Theorem:
\begin{Theorem}
\label{theo:outer_bound} For any DMS $\PXYZ$ s.t. $X \div Z \div Y$,
and a quadruplet of distortions $\bD = \{\Delta_{y,k},
\Delta_{z,k}\}_{k=1}^2$, $ \calR(\bD)_{nc} \subseteq
\calR^{**}(\bD)_{nc}.$
\end{Theorem}

The proof of this result follows the lines of the converse proof of
Theorem 1 in \cite{SMer04} and it is provided in Section
\ref{Proofs_NC}. Consider now to the case where the distortion
requirements are $\Delta_{y,1}=\infty$ and
$\Delta_{z,2}=\Delta_{z,1}$, i.e., the case where at the first stage
only Z-decoder is required to reconstruct the source and at the
second stage the Y-decoder is required to reconstruct the source
while Z-decoder is not required to improve its source reconstruction
any further. Define the degraded region $\calR(\bD)_{nc}$ of all
rates and distortions matching $\Delta_{y,1}=\infty$ and
$\Delta_{z,2}=\Delta_{z,1}$ by $\calR(\Delta_{y,2},\Delta_{z,1})$.
This special instance of our problem has been studied in
\cite{TianDiggavi07}. The outer bound obtained in
\cite{TianDiggavi07} is the following: Define the region
$\calR_{out}(\Delta_1,\Delta_2)$ to be the set of all rate pairs
$(R_{1},R_{2})$ for which there exist random variables
$(W_{1},W_{2})$ in finite alphabets $\calW_1$, $\calW_2$ s.t. the
following conditions are satisfied:
\begin{enumerate}
\item[1)]{$(W_{1},W_{2}) \div X \div Z \div Y$.}
\item[2)]{There exist deterministic maps $G_{1}: \calZ \times \calW_1 \rightarrow \tilde{X}$
and $G_{2}: \calY \times \calW_2 \rightarrow \hat{X}$ s.t.
$Ed_{z,1}(X,f_1(Z,W_1))\leq \Delta_1$ and
$Ed_{y,2}(X,f_2(Y,W_2))\leq \Delta_2$.}
\item[3)]{$|\calW_1|\leq |\calX|(|\calX|+3)+2$, $|\calW_2|\leq |\calX|+3$.}
\item[4)]{The non-negative rate vectors satisfy: \\
            $R_1 \geq I(X;W_1|Z)$, $R_1+R_2 \geq I(X;W_2|Y) +
            I(X;W_1|Z,W_2)$.}
\end{enumerate}
\begin{Theorem} \cite{TianDiggavi07}
For any discrete memoryless stochastic source with SIs under the
Markov condition $X \div Z \div Y$, $ \calR(\Delta_{1},\Delta_{2})
\subseteq \calR_{out}.$
\end{Theorem}
Note that this outer bound is straightforwardly obtainable from the
outer bound of this paper by taking $W_1 = const.$, $V = const.$,
$W_{4} = const.$ and renaming the pair $(W_{2},W_{3})$ to be
$(W_1,W_2)$ as well as setting
$(\Delta_{y,1},\Delta_{z,1},\Delta_{y,2},\Delta_{z,2})$ to be equal
$(\infty, \Delta_1, \Delta_2, \Delta_1)$, respectively, and also
disregarding $(G_{y,1}, G_{z,2})$ while renaming $(G_{z,1},
G_{y,2})$ to be $(G_1, G_2)$.

\subsubsection{Inner Bound}

Let a distortion quadruplet
$\bD\eql\{\Delta_{y,k},\Delta_{z,k}\}_{k=1}^{2}$ be given. Define
$\calR^{*}(\bD)_{nc}$ to be the set of all rate pairs
$(R_{1},R_{2})$ for which there exist RVs $\{W_{i}\}_{i=1}^{4}$ and
$V$, taking values in finite alphabets, $\{\calW_{i}\}_{i=1}^{4}$
and $\calV$, respectively, s.t. the following conditions are
satisfied:

\noindent 1. The following Markov conditions hold:
\begin{eqnarray}
\label{Markov_non-causal}
    &(W_{1},W_{2},W_{3},W_{4},V) \div X \div Z \div Y
\\
\label{Markov_non-causal111a}
    & W_{2} \div (X,W_{1},V) \div W_{3}
\end{eqnarray}
\noindent 2. There exist deterministic decoding functions
$G_{y,1}:\calY\times\calW_{1} \rightarrow\hat{\calX}$,
$G_{z,1}:\calZ\times\calW_{1}\times\calW_{2}\times\calV
\rightarrow\tilde{\calX}$, $G_{y,2}:
\calY\times\calW_{1}\times\calW_{3}\times\calV\rightarrow\check{\calX}$,
$G_{z,2}:\calZ\times\calW_{1}\times\calW_{2}\times\calW_{3}\times\calW_{4}\times\calV
\rightarrow\bar{\calX}$ such that
\begin {eqnarray}
\label{theo_1c33}
    Ed_{y,1}(X,G_{y,1}(Y,W_{1})) \leq \Delta_{y,1}
\end{eqnarray}
\begin{eqnarray}
 \label{theo_1d33}
    Ed_{z,1}(X,G_{z,1}(Z,W_{1},W_{2},V)) \leq \Delta_{z,1}
\end {eqnarray}
\begin {eqnarray}
\label{theo_1e33}
    Ed_{y,2}(X,G_{y,2}(Y,W_{1},W_{3},V)) \leq \Delta_{y,2}
\end{eqnarray}
\begin{eqnarray}
 \label{theo_1f33}
    Ed_{z,2}(X,G_{z,2}(Z,W_{1},W_{2},W_{3},W_{4},V)) \leq \Delta_{z,2}
\end {eqnarray}

\noindent 3. The alphabets $\{\calW_{k}\}_{k=1}^{4}$ and $\calV$
satisfy:
\begin{eqnarray}
\label{theo_1e}
    |\calW_{1}| \leq |\calX| + 6,
\end{eqnarray}
\begin{eqnarray}
\label{theo_1i}
    |\calV| \leq |\calX|\cdot|\calW_{1}| + 5,
\end{eqnarray}
\begin{eqnarray}
\label{theo_1f}
    |\calW_{2}| \leq |\calX|\cdot|\calW_{1}|\cdot|\calV| + 4,
\end{eqnarray}
\begin{eqnarray}
\label{theo_1g}
    |\calW_{3}| \leq
    |\calX|\cdot|\calW_{1}|\cdot|\calW_{2}|\cdot|\calV|+3,
\end{eqnarray}
\begin{eqnarray}
\label{theo_1h}
    |\calW_{4}| \leq |\calX|\cdot|\calW_{1}|\cdot|\calW_{2}|\cdot|\calW_{3}|\cdot|\calV| +
    2.
\end{eqnarray}

\noindent 4. The rates $R_{1}$ and $R_{2}$ satisfy
\begin {eqnarray}
    \label{theo_1a}
    R_{1} \geq I(X;W_{1}|Y)+I(X;W_{2},V|W_{1},Z)
\end{eqnarray}
\begin{eqnarray}
\label{theo_1b}
    R_{2}   &\geq& I(X;W_{1},V,W_{3}|Y)+I(X;W_{2}|W_{1},V,Z) \nonumber\\
            & + & I(X;W_{4}|W_{1},W_{2},W_{3},V,Z)
\end {eqnarray}

\noindent The inner bound to the rate-distortion region is
summarized in the following Theorem:
\begin{Theorem}
\label{theo:inner_bound_1} For any DMS $\PXYZ$ s.t. $X \div Z \div
Y$, and a quadruplet of distortions $\bD = \{\Delta_{y,k},
\Delta_{z,k}\}_{k=1}^2$, $\calR^{*}(\bD)_{nc} \subseteq
\calR(\bD)_{nc}$.
\end{Theorem}

The inner bound provided in this section demonstrates tradeoffs
between various schemes which are based on the notion of (strong or
weak) typicality. Recall that in the achievability schemes of
successive refinement treated in \cite{SMer04} and
\cite{TianDiggavi06} the generation of the auxiliary codebooks is
sequential: First, the codebook used at the first stage is
generated; then, for each codeword of that codebook another codebook
conditional on the codeword is generated, and so on. Every
generation of a codebook is conditioned on codewords of previously
generated codebooks. The encoder chooses the auxiliary codewords in
a sequential manner, first finding a good codeword in the first
codebook; then in the second codebook (which was generated
conditioned on that good codeword), it finds another good codeword,
and so on. The encoder proceeds until it has found all codewords
needed to describe the source at the desired accuracy at all stages
of successive refinement. The decoding process at each stage is also
performed in a sequential manner, i.e., first, the codeword in the
first codebook is found. Then, in a second codebook (matching that
codeword), a second codeword is found and so on.

When multi-group successive refinement is considered, it is unclear
if the auxiliary codebooks achieving rate-distortion bounds should
be generated ``sequentially" (in the sense described above) or ``in
parallel", with two or more codebooks generated unconditioned on one
another. The achievability scheme of this paper demonstrates a
semi-parallel approach where some of the codebooks are generated
sequentially and some in parallel. We proceed with discussing the
meaning of degraded SI at the decoders and then we briefly describe
the idea standing behind the achievability scheme.

When referring to degraded SI, the term usually used is that the
stronger Z-decoder (that has access to SI of higher quality) can do
whatever the weaker Y-decoder can do
\cite{SMer04}-\cite{MaorMerhav06}, i.e., the Z-decoder can find all
the codewords that were addressed to Y-decoder. To understand this
property, consider the following scenario: Assume that one performs
Wyner-Ziv (W-Z) coding \cite{Wyner_Ziv} for a pair $(X,Y)$ where $X$
is known at the encoder and $Y$ is known at the Y-decoder.
%\footnote{Wyner-Ziv coding is based on the following concept:
%If there is an auxiliary RV $W_1$ s.t. $W_1\div X \div Y$,
%maintaining the reconstruction constraint and the coding is
%performed in a block-wise manner, with $N$ denoting the block
%length, then the encoder can generate a codebook of a little more
%than $2^{NI(X;W_1)}$ codewords according to $P_{W1}^N$ and with high
%probability there will be at least one codeword jointly typical with
%every typical $\bX$. After generation of the codebook, the encoder
%divides it into bins (sub-groups) with each bin containing a little
%less than $2^{NI(Y;W_1)}$ codewords and each bin is assigned an
%index among a little more than $2^{N[I(X;W_{1})-I(Y;W_1)]}$ possible
%indexes. For each typical $\bX$ the encoder transmits to the decoder
%the index of a bin that contains the codeword $\bW_{1}$ that is
%jointly typical with $\bX$ and, using typical properties of
%sequences, the decoder is able to retrieve the correct codeword in
%the bin with probability tending to 1 as $N$ becomes infinitely
%large. The rate of the W-Z coding is $N[I(X;W_{1})-I(Y;W_1)] =
%NI(X;W_1|Y)$ when $N\rightarrow\infty$}.
Now, assume that the source generating the $(X,Y)$ pair is, in fact,
a ternary source, generating a triplet $(X,Y,Z)$, $X\div Z\div Y$
and that $Z$ is known at Z-decoder. Finally, assume that the W-Z
coding for Y-decoder is performed with a codebook of auxiliary
codewords generated independently of each other and each symbol of
which is generated according to $P_{W_1}$, s.t. $W_1\div X\div Z\div
Y$. Obviously, the long Markov chain also satisfies the shorter
Markov chain $W_1\div X\div Y$ required by W-Z scheme for coding for
the Y-decoder. But, due to the Markov chain $W_1\div X\div Z \div
Y$, $I(Z;W_1) \geq I(Y;W_1)$, and thus, Z-decoder is able to find
the correct codeword in the bin of size $~2^{NI(Y;W_1)}$ generated
for the Y-decoder. The question is the following - given that
Z-decoder can always find codewords addressed to the Y-decoder, how
we can exploit this property rigorously?

We interpret the degradedness of SI as follows: bins associated with
a code designed for the Z-decoder are divided into bins associated
with a code designed for the Y-decoder. Specifically, a codebook of
about $2^{NI(X;W_{1})}$ codewords is partitioned twice - first into
``large" bins of about $2^{NI(Z;W_{1})}$ codewords matching W-Z code
for the Z-decoder, and each of these bins is further partitioned
into smalle bins of about $2^{NI(Y;W_{1})}$ codewords each.% The
%inclusion of bins is possible since, as was mentioned previously,
%the Markov condition $W_{1}\div X\div Z\div Y$ guarantees that
%$I(Z;W_{1})\geq I(Y;W_{1})$.

In W-Z coding designed for communication with Y-decoder only, the
indexes of the smaller bins are directly transmitted to the
Y-decoder. Note that alternatively, one can first send to Y-decoder
an index of the larger bin and then ``refine" it with the
``internal" index of the matching small bin. %For the two-decoders
%problem, the division of codebooks is revealed to both decoders. The
%Z-decoder can decode codewords sent to the Y-decoder, while, in
%fact, Z-decoder does not need to know the ``refined index" that
%serves the Y-decoder, but it is enough for it to know the index of
%the large bin that contains the codeword chosen by the encoder.
This observation immediately leads to the following conclusion: if a
single codeword is simultaneously good for communication with both
decoders (in the sense of satisfying the reconstruction
requirements), the encoder can communicate with both decoders in a
two-stage successive manner, by first transmitting the index of a
large bin (that contains a good codeword) to both decoders (the
index is fully usable only by the Z-decoder), and then, in a
separate additional transition, sending the matching ``internal"
index which is crucial for communication with the Y-decoder (and
does not provide new information to Z-decoder). The obvious question
that arises is what happens when a single codeword is not sufficient
for communication with two decoders and more codebooks must be
created. Firstly, under certain Markov conditions, the principle of
such an hierarchical (or nested) binning can be applied as well to
conditional W-Z codes. Specifically, when the Markov condition
$(W_{1},W_{2}...,W_{i})-X-Z-Y$ holds, we obtain that
$I(Z;W_{i}|W_{1},...,W_{i-1})\geq I(Y;W_{i}|W_{1},...,W_{i-1})$.
Secondly, the real problem arises when not all codewords sent to the
Z-decoder must be revealed to the Y-decoder in the next step, and in
this case, sequential/hierarchical codesbooks generation is no
longer obviously optimal.

The coding scheme is based on the following concept: At the first
stage, three codebooks are generated, essentially, according to the
hierarchical Wyner-Ziv coding scheme. First a codebook $C_{w_1}$ of
$\sim 2^{NI(X;W_1)}$ codewords is generated according to
$P_{W_1}^N$, and is partitioned into bins of size of $\sim
2^{NI(Y;W_1)}$. Thus, there are $\sim 2^{N[I(X;W_1)-I(Y;W_1)]}$ such
bins. Due to the Markov chain $W_1\div X\div Y$,
$I(X;W_1)-I(Y;W_1)=I(X;W_1|Y)$. Next, for each $\bw_{1}\in C_{w_1}$,
a codebook $C_{v}(\bw_{1})$ of $\sim 2^{NI(X;V|W_{1})}$ codewords is
generated according to $P_{V|W_1}^N$ and is partitioned into bins of
size of $\sim 2^{NI(Z;V|W_1)}$, and each of these bins is
partitioned into smaller bins of size $\sim 2^{NI(Z;V|W_1)}$ each.
Thus, there are $\sim 2^{N[I(X;V|W_{1})-I(Z;V|W_{1})]}$ large bins
and $\sim 2^{N[I(Z;V|W_{1})-I(Y;V|W_{1})]}$ small bins within each
large bin. Due to the Markov chain (\ref{Markov_non-causal}), the
number of bins: $\sim 2^{NI(X;V|W_1,Z)}$ large bins and $\sim
2^{NI(Z;V|W_{1},Y)}$ small bins within each large one. Finally, a
codebook $C_{w_2}(\bw_{1},\bv)$ of $\sim 2^{I(X;W_2|W_1,V)}$
codewords is generated for each $\bw_{1}\in C_{w_1}$ and $\bv\in
C_{v}(\bw_1)$ according to $P_{W2|W1,V}^N$, and is partitioned into
bins of size $\sim 2^{NI(Z;W_{2}|W_{1},V)}$, so by
(\ref{Markov_non-causal}), there are $\sim
2^{NI(X;W_{2}|W_{1},V,Z)}$ such bins.

At the second stage, another two codebooks are generated - for each
$\bw_1$, $\bv(\bw_1)$ and $\bw_2(\bw_1,\bv)$, in codebook
$C_{w_3}(\bw_{1},\bv)$, the codewords are generated according to
$P_{W_3|W_1,V}^N$, and in codebook
$C_{w_4}(\bw_1,\bv,\bw_{2},\bw_{3})$, the codewords $\bW_{4}$ are
generated according to $P_{W_4|W_1,V,W_2,W_3}^N$. These codebooks
are also partitioned into bins, specifically, $C_{w_4}(\cdot)$ is
partitioned into $\sim 2^{NI(X;W_3|W_1,V,Y)}$ bins of size $\sim
2^{NI(Y;W_3|W_1,V)}$ each. Similarly, $C_5$ is partitioned into
$\sim 2^{NI(X;W_4|W_1,V,W_2,W_3,Z)}$ bins, each of them of size
$\sim 2^{NI(Z;W_4|W_1,V,W_2,W_3)}$. The key feature of this scheme
is in fact that the $C_{w_3}(\cdot)$ does not take into
consideration statistics of $\bW_{2}$. Since its codewords must yet
be used by the Z-decoder, rising typicality considerations during
the encoding/decoding process, the additional Markov condition
(\ref{Markov_non-causal111a}) is imposed on the achievability
scheme.

If the encoder succeeds to find good codewords in all five codebooks
(the details appear in the formal proof of Theorem
\ref{theo:inner_bound_1}), the rate of the first transmission,
$R_1$, is composed of three indexes of the bins that contain good
codewords in the codebooks $C_{w_1}$, $C_{v}(\cdot)$ and
$C_{w_2}(\cdot)$, where for $C_{v}(\cdot)$ only the index of the
large bin is used. In this manner, similarly to the classical W-Z
coding, only the codeword of $C_{w_1}$ serves the Y-decoder, while
all three codewords are decoded by Z-decoder. Hence, $R1 \simeq
I(X;W_1|Y) + I(X;V|W_1,Z) + I(X;W_2|W_1,V,Z)$ as is given by eq.
(\ref{theo_1a}). At the second transmission, the encoder first
refines the description of the bin of codebook $C_{v}(\cdot)$, and
then transmits the indexes of the chosen bins in codebooks
$C_{w_3}(\cdot)$ and $C_{w_4}(\cdot)$. Thus, the codewords in codes
$C_{w_1}$, $C_{v}(\cdot)$ and $C_{w_3}(\cdot)$ serve the
reconstruction in the Y-decoder and all five codewords are retrieved
correctly by Z-decoder and are used for reconstruction of the
source. The incremental rate at the second stage is $I(Z;V|W_1,Y) +
I(X;W_3|W_1,V,Y) + I(X;W_4|W_1,V,W_2,W_3)$ and therefore, the
cumulative rate at the second stage is as given by eq.
(\ref{theo_1b}).

The scheme that leads to the inner bound is interesting due to the
following: The codebook generation is not fully sequential, but some
of the codebooks are generated in parallel and are independent
(unconditioned) of each other. Unfortunately, with this approach the
rate expressions of the inner and outer bounds obtained at the
second stage are not identical and the bounds differ in additional
Markov conditions imposed on the auxiliary RVs of the direct scheme.
Yet, for the case of lossless reconstruction at either the first or
the second stage, i.e., $\Delta_{z,1}=0$ or $\Delta_{y,2}=0$,
respectively, the achievability scheme achieves communication rates
suggested by the outer bound and thus closes the gap between the
inner and the outer bounds.

\subsubsection{Special Cases}

\noindent We now confine our attention to a number of special cases
in which the gap between the outer bound and the inner bound
vanishes. First, we consider the case of distortion requirements
$\Delta_{z,1}\geq \Delta_{y,1}$ or $\Delta_{y,2}=\Delta_{y,1}$, that
is, SR with respect to only one of the decoder at either the first
or the second stages, respectively. We then consider the case of
distortion requirements $\Delta_{z,1}=0$ or  $\Delta_{y,2}=0$, that
is, lossless reconstruction at Z-decoder at the first stage, or at
the Y-decoder at the second stage, respectively. For these cases,
the achievability scheme achieves the boundary curve of the outer
bound.\vspace{12pt}
\\
\noindent \textbf{Successive Refinement}

\noindent When $\Delta_{z,1} \geq \Delta_{y,1}$ or
$\Delta_{y,2}=\Delta_{y,1,}$, the multi-decoder SR problem
degenerates to the problem of refinement of information with respect
to only one decoder at either the first or the second stage,
respectively. The requirement $\Delta_{z,1} \geq \Delta_{y,1}$ fits
the scenario where the Z-decoder performs reconstruction of the
source on the basis of the same transmission that served the
Y-decoder, so the average distortion it achieves is at least as
small as that of the Y-decoder. The requirement
$\Delta_{y,2}=\Delta_{y,1,}$ fits the scenario where the Y-decoder
is not required to refine its reconstruction at the second stage.
For these cases, the inner and the outer bounds coincide, as is
summarized in the following theorem:
\begin{Theorem}
\label{theo:SR}
    If $\Delta_{z,1} \geq \Delta_{y,1}$ or $\Delta_{y,2}=\Delta_{y,1,}$, then $\calR(\bD)_{nc}=\calR^{**}(\bD)_{nc} =
    \calR^{*}(\bD)_{nc}$. Specifically, when $\Delta_{z,1} \geq \Delta_{y,1}$,
    $\calR(\bD)_{nc}$ is given as in the Subsection \ref{Outer_Bound_def} with the rate
    inequalities replaced by
    \begin{eqnarray}
        R_{1} \geq I(X;W_{1}|Y) \textbf{ }\textbf{ }\text{ and }\textbf{ }\textbf{ }
        R_{2} \geq I(X;W_{1},W_{3}|Y) + I(X;W_{4}|W_{1},W_{3},Z),
    \end{eqnarray}
    for the auxiliary RV's satisfying $(W_{1},W_{3},W_{4}) \div X \div Z \div
    Y$.

    \noindent When $\Delta_{y,2}=\Delta_{y,1}$, $\calR(\bD)_{nc}$ is given
    as in the Subsection \ref{Outer_Bound_def} with the rate inequalities replaced by
    \begin{eqnarray}
        R_{1} \geq I(X;W_{1}|Y) + I(X;W_{2}|W_{1},Z) \textbf{ }\textbf{ }\text{ and }\textbf{ }\textbf{ }
        R_{2} \geq I(X;W_{1}|Y) + I(X;W_{2},W_{4}|W_{1},Z).
    \end{eqnarray}
    for the auxiliary RV's satisfying $(W_{1},W_{2},W_{4}) \div X \div Z \div
    Y$.
\end{Theorem}
The proof of the achievability part of Theorem \ref{theo:SR} can
easily be done by setting $W_{2}=V=\text{const.}$ for $\Delta_{z,1}
\geq \Delta_{y,1}$ and setting $W_{3}=V=\text{const.}$ for the
requirement $\Delta_{y,2}=\Delta_{y,1}$ in $\calR^*(\bD)_{nc}$. The
converse proof follows by considering a three-stage communication
scheme in the converse proof of \cite{TianDiggavi06} and combining
two of its stages into a single stage for each of the above cases.
For the case $\Delta_{z,1} \geq \Delta_{y,1}$, the first stage of
Theorem \ref{theo:SR} is essentially the first stage of
\cite{TianDiggavi06}, with the transmission addressed to the
Y-decoder. The second stage of Theorem \ref{theo:SR} is a
combination of the second and the third stages in
\cite{TianDiggavi06}, where at the second stage of
\cite{TianDiggavi06}, SR is performed with respect to the Y-decoder,
and at the third stage of \cite{TianDiggavi06}, SR is performed with
respect to the Z-decoder. For the case $\Delta_{y,2} =
\Delta_{y,1}$, the the first stage of Theorem \ref{theo:SR} matches
cumulative rates of two stages of \cite{TianDiggavi06}, there the
first stage consists of transmission of the Y-decoder and the second
stage performs SR with respect to the Z-decoder. The second stage of
Theorem \ref{theo:SR} consists of the third stage of
\cite{TianDiggavi06} with SR performed (again) with respect to the
Z-decoder. \vspace{12pt}
\\
\noindent \textbf{Lossless Reconstruction}

\noindent Consider the case of lossless reconstruction at either the
Z-decoder at the first stage or the Y-decoder at the second stage.
Similarly as in \cite{TianDiggavi07}, it turns out that in these
cases, the inner and outer bounds coincide. This observation is
summarized in the following theorem:
\begin{Theorem}
\label{theo_lossless}
    If $\Delta_{y,2}=0$ or $\Delta_{z,1}=0$, then $\calR(\bD)_{nc}=\calR^{**}(\bD)_{nc} =
    \calR^{*}(\bD)_{nc}$. Specifically, when $\Delta_{z,1}=0$,
    $\calR(\bD)_{nc}$ is given as in the Subsection \ref{Outer_Bound_def} with the rate inequalities replaced by
    \begin{eqnarray}
        R_{1} \geq I(X;W_{1}|Y) + H(X|W_{1},Z) \textbf{ }\textbf{ }\text{ and }\textbf{ }\textbf{ }
        R_{2} \geq I(X;W_{1},W_{3}|Y) + H(X|W_{1},W_{3},Z),
    \end{eqnarray}
    for the auxiliary RVs satisfying $(W_{1},W_{3}) \div X \div Z \div
    Y$.

    \noindent When $\Delta_{y,2}=0$, $\calR(\bD)_{nc}$ is given as in the Subsection \ref{Outer_Bound_def}
    with the rate inequalities replaced by
    \begin{eqnarray}
        R_{1} \geq I(X;W_{1}|Y) + I(X;W_{2}|W_{1},Z) \textbf{ }\textbf{ }\text{ and }\textbf{ }\textbf{ }
        R_{2} \geq H(X|Y),
    \end{eqnarray}
    for the auxiliary RVs satisfying $(W_{1},W_{2}) \div X \div Z \div
    Y$.
\end{Theorem}
The proof of the achievability part of Theorem \ref{theo_lossless}
can easily be done by setting $W_{4}=\text{const.}$ and $W_{2}=X$
and $V=W_{3}$ for the requirement $\Delta_{z,1}=0$ and setting
$V=W_{2}$ and $W_{3}=X$ for the requirement $\Delta_{y,2}=0$ in the
inner bound $\calR^*(\bD)_{nc}$. The converse proof follows by
applying the Heegard-Berger rate-bounds \cite{HeegardBerger85} at
both stages with the corresponding demand of lossless reconstruction
at either the first or the second stage. When the outer bound is
considered for each of the stages independently, it degenerates to
the Heegard-Berger bound and thus an intersection of the
Heegard-Berger bounds for the two stages provides a trivial outer
bound to the outer bound obtained in this paper. Since the direct
scheme achieves the communication rates suggested by the
intersection, the bounds coincide.

The key property of these special cases is the fact that not all
auxiliary RV's that determine both inner and outer bounds are active
simultaneously. Specifically, $V$, which stands for the information
transmitted to the Z-decoder at the first stage and then repeated
for the Y-decoder at the second stage, takes very specific values.
The requirement $\Delta_{z,1}=0$ means perfect reconstruction of the
source performed by the Z-decoder at the first stage. For this case,
the Z-decoder obviously needs the full information about the source,
in the spirit of Slepian-Wolf \cite{Slepian-Wolf} lossless coding.
Therefore, the optimal scheme presents the information sent to
Z-decoder at the first stage as if consisting of two (mutually
dependent) parts - information $V$ which is then revealed (refined)
to the Y-decoder at the second stage ($W_{3}=V$) and the information
needed by the Z-decoder, i.e., $W_{2}=X$. For the requirement
$\Delta_{y,2}=0$, it is expected that the Y-decoder will receive at
the second stage all the information about the source, also in the
spirit of \cite{Slepian-Wolf}. As some of this information is
already revealed to the Z-decoder at the first stage, all this
information is refined to Y-decoder at the second stage ($W_2=V$)
and then all remaining information is transmitted to Y-decoder
directly ($W_3=X$). Interestingly, the cases considered in Theorem
\ref{theo_lossless} are characterized by the same property: in both
cases, the second stage transmission serves only the weaker
Y-decoder. In the case $\Delta_{y,2}=0$, it is obvious that
$\Delta_{z,2}=0$ can be achieved as well. In the case
$\Delta_{z,1}=0$, it is trivially obtained that $\Delta_{z,2}=0$ as
well and thus, only the Y-decoder benefits from the second stage
transmission.

\section{Proofs for the Causal Case}
\label{Proofs_C}
\subsection{Proof of the Converse Part of Theorem
\ref{theo_separation}} \label{Proofs} The pure source-coding problem
is a special case of the joint source-channel problem. We provide a
proof of the converse part of Theorem \ref{theo_separation}, which
includes the converse of Theorem \ref{theo_source_causal} as a special case.
%We also
%adjust the proof of Theorem \ref{theo_separation} to show that the
%necessity part of Theorem \ref{theo_Causal_Heegard_Berger} holds.

%%---------------------------------------------------------
%%--------------------Converse Proof-----------------------
%%---------------------------------------------------------
Let $(f_{1},g_{y,1},g_{z,1},f_{2},g_{y,2},g_{z,2})$ be given encoder
and decoder functions for which the distortion constraints are
satisfied at both stages. In the proof, for the first and the second
steps of the communication protocol, we examine the mutual
information $I(\bX;\bB)$ and $I(\bX;\bar{\bB})$, respectively.

Firstly, for the case of causal state information at the encoder, we
obtain
\begin{eqnarray}
\label{Cuv_1}
    I(\bX;\bB)  & = & \sum_{i=1}^{n}I(\bX;B_{i}|B^{i-1}) \nonumber\\
                & = & \sum_{i=1}^{n}[I(\bX,B^{i-1};B_{i})-I(B^{i-1};B_{i})] \nonumber\\
                & \leq & \sum_{i=1}^{n}I(\bX,B^{i-1},S_{i+1}^n;B_{i}) \nonumber\\
                & \stackrel{(a)}{=} & \sum_{i=1}^{n}I(U_{1,i};B_{i}) \nonumber\\
                & \stackrel{(b)}{=} & n I(U_{1,T};B_{T}|T) \nonumber\\
                & \stackrel{(c)}{=} & n I(U_{1,T};B|T) \nonumber\\
                & \leq & n I(U_{1,T},T;B) \nonumber\\
                & \stackrel{(d)}{=} & n I(U_{1};B) \nonumber\\
%\end{eqnarray}
%\begin{eqnarray}
                &\stackrel{(e)}{\leq}& n C_{1},
\end{eqnarray}
where $(a)$ follows by denoting $U_{1,i}\eql
(\bX,B^{i-1},S_{i+1}^n)$ for $i\in\{1,2,...,n\}$ (note that
$U_{1,i}$ and $S_i$ are independent); (b) - by defining a time-sharing
auxiliary random variable $T$, distributed uniformly over
$\{1,2,...,n\}$ independently of all other random variables in the
system and noting that $\sum_{i=1}^nI(U_{1,i},B_i)=n\sum_{i=1}^n\frac{1}{n}I(U_{1,i},B_i)=
nI(U_{1,T},B_T|T)$;
(c) - by noting that $B = B_{T}$ since the DMC is stationary; (d) - by
denoting random variable $U_{1}\eql(U_{1,T},T)$; and finally,
(e) - by the standard channel coding theorem with causal state
information at the encoder \cite{Shannon} since $U_1 \div (A,S) \div
B$.

For non-causal availability of state information at the encoder,
note that the above defined RV's $\{U_{1,i}\}$ are, in fact, the
same RV's as these used by Gelfand and Pinsker in \cite{GP80} with
$\bX$ substituting the message $V$ of \cite{GP80}. In fact, with
$\bX$ substituting the message $V$, the converse proof of
\cite{GP80} is straightforwardly applicable to our case as all the
conditions of the proof still hold. Therefore, we can as well
upper-bound $I(\bX;\bB)$ by
\begin{eqnarray}
\label{Cuv_1a}
%    I(\bX;\bB)  & \stackrel{(a)}{=} & n[I(U_{1};B)-I(U_{1};S)] \nonumber\\
%                &\stackrel{(b)}{\leq}& n C_{1},
    I(\bX;\bB)  \stackrel{(a)}{=}  n[I(U_{1};B)-I(U_{1};S)] \stackrel{(b)}{\leq} n C_{1},
\end{eqnarray}
where $(a)$ and $(b)$ follow by \cite{GP80} and $C_{1}$ stands for
the Gel'fand-Pinsker channel capacity.

On the other hand,
\begin{eqnarray}
\label{first_rate}
    I(\bX;\bB)  & = & H(\bX) - H(\bX|\bB) \\
                & = & \sum_{i=1}^{N}[H(X_{i}|X_{1}^{i-1}) - H(X_{i}|X_{1}^{i-1},\bB)]\\
                & \stackrel{(a)}{=} & \sum_{i=1}^{N}[H(X_{i}) -
                \\\nonumber
                & - & H(X_{i}|X_{1}^{i-1},Y_{1}^{i-1},Z_{1}^{i-1},\bB)]\\
                & = &
                \sum_{i=1}^{N}I(X_{i};X_{1}^{i-1},Y_{1}^{i-1},Z_{1}^{i-1},\bB)\\
                & \stackrel{(b)}{=} & \sum_{i=1}^{N}I(X_{i};W_{1,i})\\
                & \stackrel{(c)}{=}& NI(X_{\tilde{T}};W_{1,\tilde{T}}|\tilde{T})\\
                & \stackrel{(d)}{=} & NI(X;W_{1,\tilde{T}}|\tilde{T})\\
                & = & N[I(X;W_{1,\tilde{T}},\tilde{T}) - I(X;\tilde{T})]\\
%\end{eqnarray}
%\begin{eqnarray}
                & \stackrel{(e)}{=}& NI(X;W_{1,\tilde{T}},\tilde{T})\\
                \label{last_eq1}
                & \stackrel{(f)}{=}& NI(X;W_{1}),
\end{eqnarray}
where (a) follows from the fact that the source is memoryless and
from the Markov chain $X_{i}\div (X_{1}^{i-1},\bB) \div
(Y_{1}^{i-1},Z_{1}^{i-1})$; (b) - by denoting $ W_{1,i}\eql
(X_{1}^{i-1},Y_{1}^{i-1},Z_{1}^{i-1},\bB)$; (c) - by defining a
time-sharing auxiliary random variable $\tilde{T}$, distributed
uniformly over $\{1,2,...,N\}$ independently of all other random
variables in the system; (d) - by noting that $X = X_{\tilde{T}}$
since the DMS is stationary; (e) - is again due to the fact that the
source is stationary and thus $I(X;\tilde{T})=0$; and finally, (f) -
by denoting random variable $W_{1}\eql(W_{1,\tilde{T}},\tilde{T})$.

Thus, for the first stage, we obtain $NI(X;W_{1}) \leq nC_{1}$ and
by dividing both sides of the inequality by $N$ we end up with
$I(X;W_{1}) \leq \rho_{1} C_{1}$, where $C_1$ denotes the channel
capacity for the case of causal or non-causal state availability at
the encoder. I.e, condition (\ref{theo_2a}) of Theorem
\ref{theo_separation} is satisfied.

As for the second stage, by similar considerations as in
(\ref{Cuv_1}) and (\ref{Cuv_1a}), we obtain that $I(\bX;\bar{\bB})
\leq m C_2$, where $C_2$ stands for the channel capacity of the
second channel with state information available at the encoder
(again, either causally or non-causally). Also,
\begin{eqnarray}
    I(\bX;\bar{\bB})  & = & H(\bar{\bB}) - H(\bar{\bB}|\bX)\\
                & \stackrel{(a)}{\geq}& H(\bar{\bB}|\bB) -
                H(\bar{\bB}|\bX)+I(\bB;\bar{\bB}|\bX)\\
                & = & H(\bar{\bB}|\bB)-H(\bar{\bB}|\bX,\bB)\\
                & = & I(\bX;\bar{\bB}|\bB)\\
                & = & \sum_{i=1}^{N}I(X_{i};\bar{\bB}|X_{1}^{i-1},\bB) \\
                & = & \sum_{i=1}^{N}[H(X_{i}|X_{1}^{i-1},\bB) \\
                & - & H(X_{i}|X_{1}^{i-1},\bB,\bar{\bB})]\\
                & \stackrel{(b)}{=}& \sum_{i=1}^{N}[H(X_{i}|X_{1}^{i-1},\bB,Y_{1}^{i-1},Z_{1}^{i-1}) \\
                & - & H(X_{i}|X_{1}^{i-1},\bB,\bar{\bB},Y_{1}^{i-1},Z_{1}^{i-1})]\\
                & = & \sum_{i=1}^{N}I(X_{i};\bar{\bB}|X_{1}^{i-1},\bB,Y_{1}^{i-1},Z_{1}^{i-1})\\
                \label{c}
                &  \stackrel{(c)}{=} & \sum_{i=1}^{N}I(X_{i};W_{2,i}|W_{1,i})\\
%\end{eqnarray}
%\begin{eqnarray}
                & \stackrel{(d)}{=} & NI(X;W_{2,\tilde{T}}|W_{1,\tilde{T}},\tilde{T})\\
                \label{last_eq2}
                & \stackrel{(e)}{=} & NI(X;W_{2}|W_{1})
\end{eqnarray}
where (a) follows from the fact that conditioning reduces entropy
and independence of the channels described by the following Markov
chain $(\bB, \bS)\div\bX\div(\bar{\bB},\bar{\bS})$; (b) from the
Markov chains $X_{i}\div (X_{1}^{i-1},\bB)\div
Y_{1}^{i-1}Z_{1}^{i-1}$ and $X_{i}\div
(X_{1}^{i-1},\bB,\bar{\bB})\div Y_{1}^{i-1}Z_{1}^{i-1}$; (c) come
from using the above-defined auxiliary random variables
$\{W_{1,i}\}_{i=1}^{N}$ and denoting $ W_{2,i}\eql
\bar{\bB}$;(d) comes from using the above-defined random variables
$\tilde{T}$ as well as stationarity of the
source, and finally, (e) comes from using the above defined random
variable $W_{1}$ and letting $W_{2}\eql W_{2,\tilde{T}}$. We
obtain, hence, that $NI(X;W_2|W_1) \leq mC_2$ and division of both
sides of the inequality by $N$ results in $I(X;W_2|W_1) \leq \rho_2
C_2$, which is exactly the condition (\ref{theo_2b}) of Theorem
\ref{theo_separation}.

Also, note that the Markov structure $(W_{1,i},W_{2,i})\div
X_{i}\div(Y_{i},Z_{i})$ holds for every $i=1,...,N$. Due to this
structure and the fact that the source $\PXYZ$ is stationary and
memoryless, the Markov chain $(W_{1},W_{2})\div X \div(Y,Z)$ also
holds, and thus, the condition given by (\ref{theo_1gau2}) is
satisfied.

%We pause to adjust the above proof to the Heegard-Berger setting.
%Consider a noise-free scenario and denote by $f$ the encoder
%function. Then the transmission rate is lower-bounded as follows:
%\begin{eqnarray}
%    NR  & \geq  & H(f)\\
%            & \geq  & I(\bX;f)\\
%            & =     & H(\bX) - H(\bX|f)\\
%            & =     & \sum_{i=1}^{N}[H(X_{i}) -
%                    H(X_{i}|f,X_{1}^{i-1})]\\
%            & \stackrel{(a)}{=} & \sum_{i=1}^{N}[H(X_{i}) -
%                    H(X_{i}|f,X_{1}^{i-1},Y_{1}^{i-1},Z_{1}^{i-1})]\\
%            & =     &
%            \sum_{i=1}^{N}I(X_{i};f,X_{1}^{i-1},Y_{1}^{i-1},Z_{1}^{i-1})\\
%            & \stackrel{(b)}{=} & \sum_{i=1}^{N}I(X_{i};\hat{W}_{i})\\
%            & \stackrel{(c)}{=}& NI(X_{T};\hat{W}_{T}|T)\\
%            & \stackrel{(d)}{=} & NI(X;\hat{W}|T)%\\
%\end{eqnarray}
%\begin{eqnarray}
%            & = & N[I(X;\hat{W},T) - I(X;T)]\\
%            & \stackrel{(e)}{=}& NI(X;\hat{W},T)\\
%            & \stackrel{(f)}{=}& NI(X;W),
%\end{eqnarray}
%where in (b) we define the auxiliary RVs $ \hat{W}_{i}\eql
%(X_{1}^{i-1},Y_{1}^{i-1},Z_{1}^{i-1},f)$, and (a)-(f) follow from
%the same reasons as (a)-(f) in the proof of
%(\ref{first_rate})-(\ref{last_eq1}), with some straightforward
%adjustments of RVs.

We next show that there exist functions $G_{y,1}$, $G_{z,1}$,
$G_{y,2}$ and $G_{z,2}$ that satisfy (\ref{theo_1cau2}) -
(\ref{theo_1dau4}), respectively. Denote by $g_{y,k,i}$ and
$g_{z,k,i}$ the output of the decoders Y and Z, respectively, at
stages $k=1,2$ and times $i=1,...,N$. The random variable $W_{1}$
contains $(X_{1}^{i-1},Y_{1}^{i-1},Z_{1}^{i-1},\bB)$ and $W_{2}$
contains $\bar{\bB}$. Choose the functions $G_{y,1}$, $G_{y,2}$,
$G_{z,1}$ and $G_{z,1}$ as follows:
\begin{eqnarray}
\label{def7}
    G_{y,1,\tilde{T}}(Y,W_{1}) = g_{y,1,\tilde{T}}(Y_{1}^{\tilde{T}},\bB),
\end{eqnarray}
\begin{eqnarray}
    G_{z,1,\tilde{T}}(Z,W_{1}) = g_{z,1,T}(Z_{1}^{\tilde{T}},\bB),
\end{eqnarray}
\begin{eqnarray}
    G_{y,2,\tilde{T}}(Y,W_{1},W_{2}) = g_{y,2,\tilde{T}}(Y_{1}^{\tilde{T}},\bB,\bar{\bB}),
\end{eqnarray}
\begin{eqnarray}
\label{def8}
    G_{z,2,\tilde{T}}(Z,W_{1},W_{2}) = g_{z,2,\tilde{T}}(Z_{1}^{\tilde{T}},\bB,\bar{\bB}).
\end{eqnarray}
We then have for the average distortions\footnote{The definitions in
(\ref{def7})-(\ref{def8}) determine the outputs of the decoders
functions at ``stochastic" time $\tilde{T}$. For example, the output
of the Y-decoder at the first stage at time $\tilde{T}$ is governed
by the first $\tilde{T}$ symbols of the source SI, i.e.,
$Y_{1}^{\tilde{T}}$, and the channel output $\bB$.}
\begin{eqnarray}
    Ed(X,G_{y,1}(Y,W_{1})) =
    \frac{1}{N}\sum_{i=1}^{N}Ed(X,g_{y,1,i}(Y_{1}^{i},\bB))
    \leq \Delta_{y,1},
\end{eqnarray}
\begin{eqnarray}
    Ed(X,G_{z,1}(Z,W_{1})) =
    \frac{1}{N}\sum_{i=1}^{N}Ed(X,g_{z,1,i}(Z_{1}^{i},\bB))
    \leq \Delta_{z,1},
\end{eqnarray}
\begin{eqnarray}
    Ed(X,G_{y,2}(Y,W_{1},W_{2})) &=&
    \frac{1}{N}\sum_{i=1}^{N}Ed(X,g_{y,2,i}(Y_{1}^{i}),\bB,\bar{\bB})\leq \Delta_{y,2}
\end{eqnarray}
and
\begin{eqnarray}
    Ed(X,G_{z,2}(Z,W_{1},W_{2})) &=&
    \frac{1}{N}\sum_{i=1}^{N}Ed(X,g_{z,2,i}(Z_{1}^{i}),\bB,\bar{\bB})\leq \Delta_{z,2},
\end{eqnarray}
i.e., the distortion constraints are satisfied.

%------------------------------------------------------------------
In order to complete the proof, it is left to show that the
cardinality of the alphabets of auxiliary RVs $W_{1}$ and $W_{2}$ is
limited. We use the support lemma \cite{CK81}, which is based on
Carath\'{e}odory's theorem, according to which, given $J$ real
valued continuous functionals $q_{j}$, $j=1,...,J$ on the set
$\calP(\calX)$ of probability distributions over the alphabets
$\calX$, and given any probability measure $\mu$ on the Borel
$\sigma$-algebra of $\calP(\calX)$, there exist $J$ elements
$\Q_{1},...\Q_{J}$ of $\calP(\calX)$ and $J$ non-negative reals,
$\alpha_{1},...,\alpha_{J}$, such that $\sum_{j=1}^{J}\alpha_{j}=1$
and for every $j=1,...,J$
\begin {eqnarray}
    \int_{\calP(\calX)}q_{j}(\Q)\mu(d\Q) =
    \sum_{i=1}^{J}\alpha_{i}q_{j}(\Q_{i}).
\end {eqnarray}
Before we actually apply the support lemma, we first rewrite the
relevant conditional mutual informations and the distortion
functions in a more convenient form for the use of this lemma, by
taking advantage of the Markov structures. We begin with $I(X;W_1)$:
\begin{eqnarray}
I(X;W_{1}) = H(X) - H(X|W_{1}),
\end{eqnarray}
and in the same manner, $I(X;W_2|W_1)$ becomes
\begin{eqnarray}
I(X;W_{2}|W_{1}) = H(X|W_{1}) - H(X|W_{1},W_{2}).
\end{eqnarray}

For a given joint distribution of $(X,Y,Z)$, $H(X)$ is given and
unaffected by $W_{1}$ and $W_{2}$. Therefore, in order to preserve
prescribed values of $I(X;W_1)$ and $I(X;W_2|W_1)$, it is sufficient
to preserve the associated values of $H(X|W_{1})$ and
$H(X|W_{1},W_{2})$.

We first invoke the support lemma in order to reduce the alphabet
size of $W_{1}$, while preserving the values of $H(X|W_{1})$ and
$H(X|W_{1},W_{2})$, as well as the distortions in both decoders at
both stages of communication. The alphabet of $W_{2}$ is still kept
intact at this step. Define the following functionals of a generic
distribution $Q$ over $\calX \times \calW_2$, where $\calX$ is
assumed, without loss of generality, to be $\{1,2,...,\alpha\}$,
$\alpha \eql |\calX|$:
\begin{eqnarray}
    q_{i}(Q) = \sum_{w_2}Q(x,w_2), \textrm{ } i \eql x = 1,2,...,\alpha-1,
\end{eqnarray}
\begin{eqnarray}
\label{l5}
    q_{\alpha}(Q) & = &
    -\sum_{x,w_2}Q(x,w_2)\log \sum_{w_{2}}Q(x,w_{2}),%\\
\end{eqnarray}
and
\begin{eqnarray}
\label{l6}
    q_{\alpha+1}(Q) & = &
    -\sum_{x,w_2}Q(x,w_2)\log Q(x|w_{2}).
\end{eqnarray}
Also, we define
\begin{eqnarray}
    q_{\alpha+2}(Q) =
    \sum_{y}\min_{\hat{x}}\sum_{x,w_2}Q(x,w_2)P(y|x)d_{y,1}(x,\hat{x}),
\end{eqnarray}
\begin{eqnarray}
    q_{\alpha+3}(Q) =
    \sum_{z}\min_{\tilde{x}}\sum_{x,w_2}Q(x,w_2)P(z|x)d_{z,1}(x,\tilde{x}),
\end{eqnarray}
\begin{eqnarray}
    q_{\alpha+4}(Q) =
    \sum_{y}\min_{\check{x}}\sum_{x,w_2}Q(x,w_2)P(y|x)d_{y,2}(x,\check{x})
\end{eqnarray}
and
\begin{eqnarray}
    q_{\alpha+5}(Q) =
    \sum_{z}\min_{\bar{x}}\sum_{x,w_2}Q(x,w_2)P(z|x)d_{z,2}(x,\bar{x}),
\end{eqnarray}
which along with (\ref{l5}) and (\ref{l6}) help us to preserve the
rate and distortion constraints. Applying now the support lemma for
the above defined functionals, we find that there exists a random
variable $W_{1}$ (jointly distributed with $(X,Y,Z,W_2)$, whose
alphabet size is $|W_{1}|=|\calX|+5$ and it satisfies
simultaneously:
\begin{eqnarray}
    \sum_{w_{1}}\Pr\{W_{1}=w_{1}\}q_{i}(P(\cdot|w_{1}))=P_{X}(x),
    \textrm{ } i=1,2,...,\alpha-1,
\end{eqnarray}
\begin{eqnarray}
    \sum_{w_{1}}\Pr\{W_{1}=w_{1}\}q_{\alpha}(P(\cdot|w_{1}))=H(X|W_{1}),
\end{eqnarray}
\begin{eqnarray}
    \sum_{w_{1}}\Pr\{W_{1}=w_{1}\}q_{\alpha+1}(P(\cdot|w_{1}))=H(X|W_{1},W_{2}),
\end{eqnarray}
\begin{eqnarray}
    \sum_{w_{1}}\Pr\{W_{1}=w_{1}\}q_{\alpha+2}(P(\cdot|w_{1})) =
    \min_{G_{y,1}}Ed(X,G_{y,1}(Y,W_{1})),
\end{eqnarray}
\begin{eqnarray}
    \sum_{w_{1}}\Pr\{W_{1}=w_{1}\}q_{\alpha+3}(P(\cdot|w_{1})) =
    \min_{G_{z,1}}Ed(X,G_{z,1}(Z,W_{1})),
\end{eqnarray}
\begin{eqnarray}
    \sum_{w_{1}}\Pr\{W_{1}=w_{1}\}q_{\alpha+4}(P(\cdot|w_{1})) =
    \min_{G_{y,2}}Ed(X,G_{y,2}(Y,W_{1},W_{2}))
\end{eqnarray}
and
\begin{eqnarray}
    \sum_{w_{1}}\Pr\{W_{1}=w_{1}\}q_{\alpha+5}(P(\cdot|w_{1})) =
    \min_{G_{z,2}}Ed(X,G_{z,2}(Z,W_{1},W_{2})).
\end{eqnarray}
Having found a random variable $W_{1}$, we now proceed to reduce the
alphabet of $W_{2}$ in a similar manner, where this time, we have
$\beta = |\calX|\cdot|\calW_{1}|-1$ constraints to preserve the
joint distribution of $(X,W_{1})$, just defined, and $3$ more
constraints to preserve the second-stage rate and distortions.
Applying the support lemma, we obtain that $W_{2}$ satisfies all the
desired rate-distortion constraints and the necessary alphabet size
of $W_{2}$ is upper-bounded by
\begin{eqnarray}
    |\calW_{2}| \leq |\calX|\cdot|\calW_{1}|+2.
\end{eqnarray}
This completes the proof of the converse part of Theorem
\ref{theo_separation}.

\subsection{Proof of the Direct Part of Theorem \ref{theo_source_causal}}
\label{DirectProof}

%\subsection{Proof of Theorem \ref{theo_source_causal}}
Let $W_{1}$, $W_{2}$, $G_{y,1}$, $G_{y,2}$, $G_{z,1}$ and $G_{z,2}$
be some elements in the definition of $\calR^*(\bD)_c$ that achieve
a given point in that region. We next describe the mechanisms of
random code selection and the encoding and decoding operations.

\vspace{0.5\baselineskip} \noindent \textsl{Code Generation}:

\noindent Let $\epsilon_1>0$, $\epsilon_2>0$ and $\delta>0$ be
arbitrary small and select $R_1 \geq I(X;W_1)+\epsilon_1+\delta$ and
$\Delta_{R} \eql R_{2}-R_{1}$, $\Delta_{R} \geq
I(X;W_{2}|W_{1})+\epsilon_{2}+\delta$. For the first stage,
$2^{NR_{1}}$, sequences of length $N$, $\{\bW_{1}(k)\}$,
$k\in[1,...,2^{NR_{1}}]$, are drawn independently from $\Tgwa$. Let
us denote the set of these sequences by $\calC_{1}$. For each
codeword $\bW_{1}(k) = \bw_1$, a set of $2^{N\Delta_{R}}$
second-stage codewords $\{\bW_{2}(k,j)\}$, $j \in
[1,...,2^{NR_{2}}]$, are independently drawn from $\Tgwbwa$. We
denote this set by $\calC_{2}(k)$ and its elements by
$\{\bW_{2}(k,j)\}$. Note that the $2^{NR_{1}}$ sets
$\{\calC_{2}(\cdot)\}$ may not be all mutually exclusive.

\vspace{0.5\baselineskip} \noindent \textsl{Encoding}:

\noindent Upon receiving a source sequence $\bx$, the encoder acts
as follows:
\begin {enumerate}
\item {If $\bx \in \Tgx$ and the codebook $\calC_{1}$ contains a sequence
    $\bW_{1}(k)=\bw_{1}$ s.t. the pair $(\bx,\bw_{1}) \in \Tgxwa$, the fist such
    index $k$ is chosen for transmission at the first stage. Next, if the codebook
    $\calC_{2}(k)$ contains a sequence $\bW_{2}(k,j)=\bw_{2}$
    s.t. $(\bx,\bw_{1},\bw_{2}) \in \Tgxwavwb$,  the first such index $j$ is chosen
    for transmission at the second stage.
    }
\item {If $\bx \notin \Tgx$, or $\not\exists \bW_{1}(k)=\bw_{1} \textrm{ }
    \textrm{ s.t. } (\bx,\bw_{1}) \in \Tgxwa$, or $\not\exists
    \bW_{2}(k,j)=\bw_{2}
    \textrm{ }\textrm{ s.t. } (\bx,\bw_{1},\bw_{2}) \in \Tgxwavwb$,
    an arbitrary error message is transmitted at both stages.
}
\end {enumerate}

\vspace{0.5\baselineskip} \noindent \textsl{Decoding}:

\noindent The decoders of the first stage retrieves the first-stage
codeword according to its index and generates the reproduction by
$\hat{X}_{i}=G_{y,1}\left(Y_{i},W_{1,i}(k)\right)$ and
$\tilde{X}_{i}=G_{z,1}\left(Z_{i},W_{1,i}(k)\right)$,
$i\in[1,2,...,N]$. Similarly, the decoders of the second stage
retrieve both the first-stage and the second-stage codewords and
creates the reconstruction of the source according to
$\check{X}_{i}=G_{y,2}\left(Y_{i},W_{1,i}\left(k\right),W_{2,i}\left(k,j\right)\right)$
and
$\bar{X}_{i}=G_{z,2}\left(Z_{i},W_{1,i}\left(k\right),W_{2,i}\left(k,j\right)\right)$,
$i\in[1,2,...,N]$.

We now turn to the analysis of the error probability and the
distortions. For each $\bx$ and a particular choice of codes
$\calC_{1}$ and $\{\calC_{2}(\cdot)\}$, the possible causes for
error message are:

\begin{enumerate}
  \item {$\bx \notin \Tgx$. Let the probability of this event be defined as $P_{e_{1}}$.
  }
  \item {$\bx \in \Tgx$, but in the codebook $\calC_{1}$ $\not\exists \bw_{1}$ s.t. $(\bx,\bw_{1}) \in \Tgxwa$.
    Let the probability of this event be defined as $P_{e_{2}}$.
  }
  \item {$\bx \in \Tgx$, and the codebook $\calC_{1}$ contains $\bw_{1}$ s.t. $(\bx,\bw_{1})
    \in \Tgxwa$, but $\not\exists \bw_{2} \in \calC_{2}(\bw_{1})$ s.t. $(\bx,\bw_{1},\bw_{2}) \in \Tgxwavwb$.
    Let the probability of this event be defined as $P_{e_{3}}$.
  }
\end{enumerate}

Note that if none of those events occur, then, for the sufficiently
large $N$, by the Markov Lemma \cite[pp.\ 436, Lemma 14.8.1]{CT91}
applied twice, the following is satisfied at both stages: with high
probability $(\bX,\bZ,\mathbf{\hat{X}})\in\Tgxzhx$ and
$(\bX,\bY,\mathbf{\tilde{X}})\in\Tgxytx$. In particular, the first
application of the Markov Lemma occurs due to the Markov chain
$(W_{1},W_{2})\div X\div (Y,Z)$: Note that by the way of creation,
$\bX$, $\bY$ and $\bZ$ are jointly typical with high probability and
also, with high probability, $\bX$, $\bW_1$ and $\bW_2$ are jointly
typical. Therefore, by the Markov Lemma, $(\bX,\bY,\bZ,\bW_1,\bW_2)$
are also jointly typical with high probability. Also, note that due
to the fact that the source is memoryless and by the way of creation
of the reconstructions, the following Markov chains hold: $\bX\div
(\bY,\bW_1) \div \mathbf{\hat{X}}$ and $\bX\div (\bZ,\bW_1) \div
\mathbf{\tilde{X}}$, and also, at the second stage, $\bX\div
(\bY,\bW_1,\bW_{2}) \div \mathbf{\check{X}}$ and $\bX\div
(\bZ,\bW_1,\bW_{2}) \div \mathbf{\bar{X}}$. By the second
application of the Markov Lemma, we obtain that with high
probability $\bX$ is jointly typical with $\mathbf{\hat{X}}$ and
$\mathbf{\tilde{X}}$ at the first stage and with
$\mathbf{\check{X}}$ and $\mathbf{\bar{X}}$ at the refinement stage.
The probability that one or more of the above typicality relations
do not hold vanishes as $N$ becomes infinitely large. The joint
typicality of $(\bX,\mathbf{\hat{X}})$, $(\bX,\mathbf{\tilde{X}})$,
$(\bX,\mathbf{\check{X}})$ and $(\bX,\mathbf{\bar{X}})$ imposes that
the distortion constraints (\ref{theo_1cau2})-(\ref{theo_1dau4}) are
satisfied when $N$ is large enough.

It remains to show that the probability of sending an error message, $P_{e}$,
vanishes when $N$ is large enough. $P_{e}$ is bounded by
\begin {eqnarray}
\label{Pe}
    P_{e} \leq
    P_{e_{1}}+P_{e_{2}}+P_{e_{3}}.
\end {eqnarray}

\noindent The fact that $P_{e_{1}}\rightarrow 0$ follows from the
properties of typical sequences \cite{CT91}. As for $P_{e_{2}}$, we
have:
\begin {equation}
\label{P_e2a}
    P_{e_{2}} \eql \prod_{k=1}^{2^{NR_{1}}}\Pr\left\{\left(\bx,\bW_{1}(k)\right) \notin \Tgxwa\right\}.
\end {equation}
Now, for every $k$:
\begin {eqnarray}
\label{P_e2i}
    \Pr\left\{\left(\bx,\bW_{1}(k)\right) \notin \Tgxwa\right\} & = & 1 -
        \Pr\left\{\left(\bx,\bW_{1}(k)\right) \in \Tgxwa\right\} \\
    & = & 1 - \frac{|\Tgxwa|}{|\Tgwa||\Tgx|} \nonumber\\
    & \leq &1 - 2^{-N[I(X;W_{1})+\epsilon_{1}]} \nonumber,
\end {eqnarray}
where the last equation follows from the size of typical sequences
as are given in \cite{CT91}. Substitution of (\ref{P_e2i}) into
(\ref{P_e2a}) and application of the well-known inequality
$(1-v)^{N}\leq\exp(-vN)$, provides us with the following upper-bound
for $N\rightarrow\infty$:
\begin {equation}
    P_{e_{2}} \leq \Big[1 - 2^{-N[I(X;W_{1})+\epsilon_{1}]}\Big]^{2^{nR_1}} \leq
        \exp\left\{-2^{NR_{1}}\cdot2^{-N[I(X;W_1)+\epsilon_{1}]}\right\} \rightarrow 0,
\end {equation}
double-exponentially rapidly since $R_1 \geq I(X;W_{1})+\epsilon_{1}
+ \delta$.

\indent To estimate $P_{e_{3}}$, we repeat the technique of the
previous step:
\begin {equation}
\label{P_e3a}
    P_{e_{3}} \eql \prod_{j=1}^{2^{NR_{2}}}\Pr\left\{\left(\bx,\bw_{1},\bW_{2}(\bw_{1},j)\right) \notin \Tgxwavwb\right\}.
\end {equation}
Again, by the property of the typical sequences, for every $j$:
\begin {eqnarray}
\label{P_e3Cu}
    \Pr\left\{\left(\bx,\bw_{1},\bW_{2}(\bw_{1},j)\right) \notin \Tgxwavwb\right\} \leq 1 - 2^{-N[I(X;W_{2}|W_{1})+\epsilon_{2}]},
\end {eqnarray}
and therefore, substitution of (\ref{P_e3Cu}) into (\ref{P_e3a})
gives
\begin {equation}
    P_{e_{3}} \leq \Big[1 - 2^{-N[I(X;W_{2}|W_{1})+\epsilon_{2}]}\Big]^{2^{NR_{2}}} \leq
        \exp\left\{-2^{NR_{2}}\cdot2^{-N[I(X;W_{2}|W_{1})+\epsilon_{2}]}\right\} \rightarrow 0,
\end {equation}
double-exponentially rapidly since $R_{2} \geq
I(X;W_{2}|W_{1})+\epsilon_{2} + \delta$.

\indent Since $P_{e_{i}}\rightarrow 0$ for $i=1$,$2$,$3$, their sum
tends to zero as well, implying that there exist at least one choice
of a codebook $\calC_{1}$ and related choices of sets
\{$\calC_{2}(\cdot)$\}, that give rise to the reliable source
reconstruction at both stages with communication rates $R_{1}$ and
$\Delta_{R} = R_{2} - R_{1}$.

\section{Proofs for the Non-Causal Case}
\label{Proofs_NC}

\subsection{Outer Bound} \label{Outer_Bound} The proof of the outer
bound follows the lines of the proof of Theorem 1 in \cite{SMer04}.
Assume that we have an
$(n,M_1,M_2,\{\Delta_{y,k},(\Delta_{z,k}\}_{k=1}^2))$ SR code for
the source $X$ with SI $(Y,Z)$, as in Definition~\ref{def3}. We will
show the existence of a quintuplet $(W_{1},W_{2},W_{3},W_{4},V)$
that satisfies the conditions 1--4 in the definition of
$\calR^{**}(\bD)_{nc}$. First, note that
\begin{eqnarray}
NR_1 &\geq& H(f_1)\geq I(\bX;f_1|\bY) = I(\bX;f_1,\bZ|\bY) -
            I(\bX;\bZ|f_1,\bY)\nonumber\\
  &=& \sum_{i=1}^n \left[
      I(X_i;f_1,\bZ|X^{i-1},\bY)-I(\bX;Z_i|f_1,\bY,Z^{i-1})
        \right].
    \label{eq:conv1}
\end{eqnarray}
For notational convenience, we denote $Z^{i-1}Z_{i+1}^N =
Z^{N\backslash i}$, and use a similar notation for $X$ and $Y$.
Since $(X_i,Y_i)$ and $(X^{i-1},Y^{N\backslash i})$ are independent,
we have, for the first term in the summand of ~(\ref{eq:conv1}):
\begin{eqnarray}
 I(X_i;f_1,\bZ|X^{i-1},\bY) &=& H(X_i|Y_i,X^{i-1},Y^{N\backslash i})
       -H(X_i|Y_i,X^{i-1},Y^{N\backslash i},f_1,\bZ)\nonumber\\  &=& H(X_i|Y_i) - H(X_i|Y_i,X^{i-1}Y^{N\backslash i},f_1,\bZ)
     \nonumber\\
 &=& I(X_i;X^{i-1},Y^{N\backslash i},f_1,\bZ|Y_i).\label{eq:conv2}
\end{eqnarray}

Next, due to the Markov structure
\begin{eqnarray} Z_i\div
(X_i,Y_i)\div (X^{N\backslash i},f_1,Z^{i-1}, Y^{N\backslash
i})\label{eq:conv2.1}
\end{eqnarray}
we have, for the second term in the summand of ~(\ref{eq:conv1}):
\begin{eqnarray}
    I(\bX;Z_i|f_1,\bY,Z^{i-1}) &=& H(Z_i|f_1,\bY,Z^{i-1}) -
        H(Z_i|\bX, f_1,\bY,Z^{i-1})\nonumber\\
    &=& H(Z_i|f_1,\bY,Z^{i-1}) -H(Z_i|X_i, f_1,\bY,Z^{i-1}) \nonumber\\
    &=& I(X_i;Z_i|f_1,\bY,Z^{i-1}).\label{eq:conv3}
\end{eqnarray}
Substituting~(\ref{eq:conv2}) and~(\ref{eq:conv3})
in~(\ref{eq:conv1}), we obtain
\begin{eqnarray}
    NR_1 &\geq& \sum_{i=1}^N \left[
        I(X_i;X^{i-1},Y^{N\backslash i},f_1,\bZ|Y_i)
        - I(X_i;Z_i|f_1,\bY,Z^{i-1}) \right] \nonumber\\
        &=&  \sum_{i=1}^N \left[
             I(X_i;Y^{N\backslash i},f_1,Z^{i-1}|Y_i) +
             I(X_i;X^{i-1},Z_i^N|Y_i, f_1,Y^{N\backslash i}, Z^{i-1})
             - I(X_i;Z_i|f_1,\bY,Z^{i-1}) \right] \nonumber\\
        &=& \sum_{i=1}^n \left[
            I(X_i;f_1,Y^{N\backslash i},Z^{i-1}|Y_i) +
            I(X_i;X^{i-1},Z_{i+1}^N|Y_i,Z_i,f_1,Y^{N\backslash i}, Z^{i-1})
    \right].
   \label{eq:conv4}
\end{eqnarray}
The Markovity of $X\div Z\div Y$ implies
\begin{eqnarray}
    Y_i\div Z_i\div (X_i,f_1,Y^{N\backslash i},Z^{i-1}),
         \label{eq:conv5}
\end{eqnarray}
and we have for the second term in~(\ref{eq:conv4})
\begin{eqnarray}
\lefteqn{
    I(X_i;X^{i-1},Z_{i+1}^N|f_1,\bY,Z^i) }\nonumber\\
    &=& H(X_i|f_1,\bY,Z^i) - H(X_i|f_1,\bY,\bZ,X^{i-1})\nonumber\\
    &=& H(X_i,Y_i|f_1,Y^{N\backslash i},Z^i) - H(Y_i|f_1,Y^{N\backslash i},Z^i)
     - H(X_i|f_1,\bY,\bZ,X^{i-1})\nonumber\\
    &=& H(Y_i|X_i,f_1,Y^{N\backslash i},Z^i)
     +  H(X_i|f_1,Y^{N\backslash i},Z^i)
     - H(Y_i|f_1,Y^{N\backslash i},Z^i)
     - H(X_i|f_1,\bY,\bZ,X^{i-1})\nonumber\\
    &\stackrel{(a)}{=}&
     H(X_i|f_1,Y^{N\backslash i},Z^i)
     - H(X_i|f_1,\bY,\bZ,X^{i-1}) \nonumber \\
    & = & I(X_i;Y_i,Z_{i+1}^N,X^{i-1}|f_1,Y^{N\backslash i},Z^i)\nonumber\\
    &\stackrel{(b)}{=}& I(X_i;Z_{i+1}^N,X^{i-1}|f_1,Y^{N\backslash i},Z^i)
       \label{eq:conv6}
\end{eqnarray}
where in $(a)$ was used the Markov chain $X_{i}\div
(f_1,Y^{N\backslash i},Z^i) \div Y_{i}$. To justify (b), note that
$f_{1}$ is a function of $\bX$ and due to this feature, the fact
that the source is a DMS and the Markov condition $X\div Z \div Y$,
we obtain that $X_{i}\div (f_{1},Y^{N\backslash i},\bZ,X^{i-1})\div
Y_{i}$.

Substituting~(\ref{eq:conv6}) in~(\ref{eq:conv4}), we get
\begin{eqnarray}
    \label{eq:conv7}
    NR_1 & \geq & \sum_{i=1}^N \left[
        I(X_i;f_1,Y^{N\backslash i},Z^{i-1}|Y_i)
        +  I(X_i;Z_{i+1}^N,X^{i-1}|f_1,Y^{N\backslash i},Z^i)\right]
        \\
        & \geq & \sum_{i=1}^N \left[
        I(X_i;f_1,Y^{N\backslash i},Z^{i-1}|Y_i)
        +  I(X_i;Z_{i+1}^N|f_1,Y^{N\backslash i},Z^i)\right]
                                              \label{eq:conv7new}\\
        & \stackrel{(a)}{=} & \sum_{i=1}^N \left[
        I(X_i;f_1,Y^{N\backslash i}|Y_i)
        + I(X_i;Z^{i-1}|f_{1},\bY) + I(Z_i;Z^{i-1}|f_1,\bY,X_{i})
        +  I(X_i;Z_{i+1}^N|f_1,Y^{N\backslash i},Z^i)\right]\nonumber\\
        & = & \sum_{i=1}^N \left[
        I(X_i;f_1,Y^{N\backslash i}|Y_i)
        + I(X_i,Z_{i};Z^{i-1}|f_{1},\bY)
        +  I(X_i;Z_{i+1}^N|f_1,Y^{N\backslash i},Z^i)\right]\nonumber\\
        & = & \sum_{i=1}^N \left[
        I(X_i;f_1,Y^{N\backslash i}|Y_i)
        + I(Z_{i};Z^{i-1}|f_{1},\bY)
        + I(X_{i};Z^{i-1}|f_{1},\bY,Z_{i})
        +  I(X_i;Z_{i+1}^N|f_1,Y^{N\backslash i},Z^i)\right]\nonumber\\
        & \geq & \sum_{i=1}^N \left[
        I(X_i;f_1,Y^{N\backslash i}|Y_i)
        + I(X_{i};Z^{i-1}|f_{1},\bY,Z_{i})
        +  I(X_i;Z_{i+1}^N|f_1,Y^{N\backslash i},Z^i)\right]\nonumber\\
        & \stackrel{(b)}{=} & \sum_{i=1}^N \left[
        I(X_i;f_1,Y^{N\backslash i}|Y_i)
        + I(X_{i};Z^{i-1}|f_{1},Y^{N\backslash i},Z_{i})
        +  I(X_i;Z_{i+1}^N|f_1,Y^{N\backslash i},Z^i)\right]\nonumber\\
        & = & \sum_{i=1}^N \left[
        I(X_i;f_1,Y^{N\backslash i}|Y_i)
        +  I(X_i;Z^{N\backslash i}|f_1,Y^{N\backslash
i},Z_i)\right], \label{eq:conv7new1}
\end{eqnarray}
where (a) is due to the Markov relation $Z_{i}\div
(f_{1},\bY,X_{i})\div Z^{i-1}$ and (b) is due to the Markov chain
$Y_{i}\div Z_i \div (f_{1},Y^{N\backslash i},Z^{i-1},X_{i})$ that
implies $Y_{i}\div (f_{1},Y^{N\backslash i},Z_{i})\div Z^{i-1}$ and
$Y_{i}\div (f_{1},Y^{N\backslash i},Z_{i},X_{i})\div Z^{i-1}$.

Before defining the auxiliary random variables, we bound $R_{2}$
from below. We do that by repeating the steps
(\ref{eq:conv1})-(\ref{eq:conv7}) of lower-bounding $R_{1}$ with a
pair $(f_{1},f_{2})$ substituting $f_{1}$ in each step:
\begin{eqnarray}
    NR_2 &\geq& H(f_1,f_2)\geq H(\bX;f_1,f_2|\bY)
         \geq I(\bX;f_{1},f_2,\bZ|\bY)-I(\bX;\bZ|f_{1},f_{2},\bY)\nonumber\\
         & \geq & \sum_{i=1}^N\left[
        I(X_i;f_1,f_{2},Y^{N\backslash i},Z^{i-1}|Y_i)
        +  I(X_i;Z_{i+1}^N,X^{i-1}|f_1,f_{2},Y^{N\backslash
i},Z^i)\right]\label{eq:conv8}
\end{eqnarray}
Define the random variables $W_{1,i} = (f_1,Y^{N\backslash i})$,
$V_{i} = Z^{i-1}$, $W_{2,i} = Z_{i+1}^N $, $W_{3,i} = f_2$ and
$W_{4,i} = X^{i-1}$.
%\begin{eqnarray}
%    &&W_{1,i} = (f_1,Y^{N\backslash i})\label{eq:conv9W1}\\
%    &&V_{i} = Z^{i-1}\label{eq:conv9V}\\
%    &&W_{2,i} = Z_{i+1}^N \label{eq:conv9W2}\\
%    &&W_{3,i} = f_2\label{eq:conv9W3} \\
%    &&W_{4,i} = X^{i-1}\label{eq:conv9W4}
%\end{eqnarray}
With these definitions \footnote{Note that different choices of
auxiliary RVs are possible. For example, one may choose: $W_{1,i} =
f_1,Y^{N\backslash i}$, $V_{i} = (W_{1,i},Z^{i-1})$, $W_{2,i} =
(V_{i},Z_{i+1}^N)$, $W_{3,i}=(V_{i},f_2)$,
$W_{4,i}=(W_{2,i},W_{3,i},X^{i-1})$. This choice would result in the
following Markov chain: $W_{1,i} \div V_{i} \div (W_{2,i},W_{3,i})
\div W_{4,i} \div X_i \div Z_i \div Y_i $.}, we have the Markov
structure
\begin{eqnarray} (W_{1,i},W_{2,i},W_{3,i},W_{4,i},V_i)\div X_i\div Z_i\div Y_i
   \label{eq:convMarkov}
\end{eqnarray}
and the bounds~(\ref{eq:conv7new1}) and~(\ref{eq:conv8}) become
%\begin{eqnarray} R_1 &\geq&
%                \frac{1}{N}\sum_{i=1}^N \left[ I(X_i;W_{1,i}|Y_i)
%                + I(X_i;W_{2,i}|W_{1,i},Z_i) \right]
%  \label{eq:convB1}\\
%                R_2 &\geq& \frac{1}{N}\sum_{i=1}^N \left[ I(X_i;W_{1,i},W_{3,i}|Y_i) +
%                I(X_i;W_{2,i},W_{4,i}|W_{1,i},W_{3,i},Z_i) \right].\label{eq:convB2}
%\end{eqnarray}
\begin{eqnarray} R_1 &\geq&
                \frac{1}{N}\sum_{i=1}^N \left[ I(X_i;W_{1,i}|Y_i)
                + I(X_i;W_{2,i},V_{i}|W_{1,i},Z_i) \right]
  \label{eq:convB1}\\
                R_2 &\geq& \frac{1}{N}\sum_{i=1}^N \left[ I(X_i;W_{1,i},V_{i},W_{3,i}|Y_i) +
                I(X_i;W_{2,i},W_{4,i}|W_{1,i},W_{3,i},V_{i},Z_i) \right].\label{eq:convB2}
\end{eqnarray}
Let $J$ be a random variable, independent of $X$, $Y$, and $Z$, and
uniformly distributed over the set $\{1,2,\ldots,N\}$. Define the
random variables $W_{1}=(J,W_{1,J})$, $V=(J,V_{J})$,
$W_{2}=(J,W_{2,J})$, $W_{3}=(J,W_{3,J})$ and $W_{4}=(J,W_{4,J})$.
The Markov relations~(\ref{eq:convMarkov}) still hold, that is
\begin{eqnarray}
    (W_{1},W_{2},W_{3},W_{4},V) \div X\div Z\div Y,
    \label{eq:convMarkov2}
\end{eqnarray}
and therefore the condition 1 in the definition of
$\calR^{**}(\bD)_{nc}$ is satisfied.

We proceed to show the existence of functions $G_{y,1}$, $G_{z,1}$,
$G_{y,2}$ and $G_{z,2}$ satisfying the second condition. Denote by
$g_{y,k,l}$ and $g_{z,k,l}$ the output of the Y and Z decoders,
respectively, at iteration $k$ and time $l$, $k=1,2$, $1\leq l\leq
N$. The random variable $W_{1}$ contains $f_{1} Y^{N\backslash J}$.
At the same time, the triplet $(W_{1},V,W_{2})$ contains $f_1
Z^{N\backslash J}$ and so on. Therefore, let us choose the functions
$G_{y,1}$, $G_{z,1}$, $G_{y,2}$ and $G_{z,2}$ as follows
\begin{eqnarray}
G_{y,1,J}(Y,W_{1}) &=& g_{y,1,J}(\bY,f_1)\label{eq:conv13}\\
G_{z,1,J}(Z,W_{1},W_{2},V) &=&
g_{z,1,J}(\bZ,f_1).\label{eq:conv14}\\
G_{y,2,J}(Y,W_{1},W_{3},V) &=& g_{y,2,J}(\bY,f_1,f_2)\label{eq:conv15}\\
G_{z,2,J}(Z,W_{1},W_{2},W_{3},W_{4},V) &=&
g_{z,2,J}(\bZ,f_1,f_2).\label{eq:conv16}
\end{eqnarray}
Then, for the distortions we have
\begin{eqnarray}
    \mathbb{E}d_{y,1}(X,G_{y,1}(Y,W_{1}))
        &=& \frac{1}{N}\sum_{j=1}^N
        \mathbb{E} d_{y,1}(X,g_{y,1,j}(\bY,f_1))\leq \Delta_{y,1}\label{eq:convd1}\\
    \mathbb{E} d_{z,1}(X,G_{z,1}(Z,W_{1},W_{2},V))
        &=& \frac{1}{N}\sum_{j=1}^N
        \mathbb{E} d_{z,1}(X,g_{z,1,j}(\bZ,f_1)) \leq \Delta_{z,1}\label{eq:convd2}\\
    \mathbb{E}d_{y,2}(X,G_{y,2}(Y,W_{1},W_{3},V))
        &=& \frac{1}{N}\sum_{j=1}^N
        \mathbb{E} d_{y,2}(X,g_{y,2,j}(\bY,f_1,f_2))\leq \Delta_{y,2}\nonumber\\
        \label{eq:convd3}\\
    \mathbb{E} d_{z,2}(X,G_{z,2}(Z,W_{1},W_{2},W_{3},W_{4},V))
        &=& \frac{1}{N}\sum_{j=1}^N
        \mathbb{E} d_{z,2}(X,g_{z,2,j}(\bZ,f_1,f_2)) \leq \Delta_{z,2}\nonumber\\\label{eq:convd4}
\end{eqnarray}
Hence, condition 2 in the definition of $\calR^{**}(\bD)_{nc}$ is
satisfied.

To prove that condition 4 of that definition holds, we have to show
that the bounds~(\ref{theo_1a3aa}) and~(\ref{theo_1b3aa}) can be
written in a single letter form with $W_{1}$, $W_{2}$, $W_{3}$ and
$W_{4}$. The following chain of equalities holds
\begin{eqnarray}
I(X;W_{1}|Y) &=& H(W_{1}|Y)- H(W_{1}|X,Y) \nonumber\\
             &=& H(W_{1}|Y)-H(W_{1}|X)\nonumber\\
             &=& I(W_{1};X)-I(W_{1};Y)\nonumber\\ &=& H(X) - H(X|W_{1}) - H(Y)
+H(Y|W_{1})\nonumber\\ &=& H(X) - H(X|J,W_{1,J})-H(Y) +H(Y|J,W_{1,J})\nonumber\\
%\end{eqnarray}
%\begin{eqnarray}
&=& \frac{1}{N}\sum_{i=1}^N H(X_i)- \frac{1}{N}\sum_{i=1}^N
H(X_i|W_{1,i})
     -\frac{1}{N}\sum_{i=1}^N H(Y_i)  +  \frac{1}{N}\sum_{i=1}^N H(Y_i|W_{1,i})
       \nonumber\\
&=& \frac{1}{N}\sum_{i=1}^N I(X_i;W_{1,i}|Y_i) \label{eq:conv19}
\end{eqnarray}
where the last equality is due to~(\ref{eq:convMarkov}). In a
similar manner, we get
\begin{eqnarray}
    I(X;W_{2},V|W_{1},Z) &=& I(X;J,W_{2,J},J,V_{J}|J,W_{1,J},Z) = I(X;W_{2,J},V_{J}|J,W_{1,J},Z)\nonumber\\
    &=& H(X|J,W_{1,J},Z) - H(X|J,W_{1,J},W_{2,J},V_{J},Z)\nonumber\\
    &=& \frac{1}{N}\sum_{i=1}^N H(X_i|i,W_{1,i},Z_i)
        -\frac{1}{N}\sum_{i=1}^N H(X_i|i,W_{1,i},W_{2,i},V_{i}Z_i)\nonumber\\
    &=& \frac{1}{N}\sum_{i=1}^N I(X_i;W_{2,i},V_{i}|W_{1,i},Z_i). \label{eq:conv20}
\end{eqnarray}
In view of~(\ref{eq:conv19}), (\ref{eq:conv20}), the
bound~(\ref{eq:convB1}) can be written as
\begin{eqnarray}
    R_1\geq I(X;W_{1}|Y) + I(X;W_{2},V|W_{1},Z). \label{eq:convB1final}
\end{eqnarray}
In a similar manner, we shown that~(\ref{eq:convB2}) can be written
as \begin{eqnarray} R_2 \geq
I(X;W_{1},W_{3},V|Y)+I(X;W_{2},W_{4}|W_{1},W_{3},V,Z).
\label{eq:convB2final}
\end{eqnarray}
Specifically,
\begin{eqnarray}
I(X;W_{1},W_{3},V|Y) &=& H(W_{1},W_{3},V|Y)- H(W_{1},W_{3},V|X,Y)
\nonumber \\
&=&H(W_{1},W_{3},V|Y)-H(W_{1},W_{3},V|X)\nonumber\\
&=& H(W_{1},W_{3},V)-H(W_{1},W_{3},V|X) \nonumber \\
&-&(H(W_{1},W_{3},V)-H(W_{1},W_{3},V|Y))\nonumber\\
&=& I(W_{1},W_{3},V;X)-I(W_{1},W_{3},V;Y)\nonumber\\
&=& H(X) - H(X|W_{1},W_{3},V) - H(Y)\nonumber\\
&+& H(Y|W_{1},W_{3},V)\nonumber\\
&=& H(X) - H(X|J,W_{1,J},J,W_{3,J},J,V_{J})-H(Y) \nonumber\\
&+& H(Y|J,W_{1,J},J,W_{3,J},J,V_{J})\nonumber\\
&=& \frac{1}{N}\sum_{i=1}^N H(X_i)- \frac{1}{N}\sum_{i=1}^N
H(X_i|W_{1,i},W_{3,i},V_{i})
     -\frac{1}{N}\sum_{i=1}^N H(Y_i) \nonumber\\
%\end{eqnarray}
%\begin{eqnarray}
& + & \frac{1}{N}\sum_{i=1}^N H(Y_i|W_{1,i},W_{3,i},V_{i})
       \nonumber\\
&=& \frac{1}{N}\sum_{i=1}^N I(X_i;W_{1,i},W_{3,i},V_{i}|Y_i)
\label{eq:conv19a}
\end{eqnarray}
where the last equality is due to~(\ref{eq:convMarkov}). In a
similar manner, we get
\begin{eqnarray}
    I(X;W_{2},W_{4}|W_{1},W_{3},V,Z) &=& I(X;J,W_{2,J},J,W_{4,J}|J,W_{1,J},J,W_{3,J},V,Z) \nonumber\\
    & = & I(X;W_{2,J},W_{4,J}|J,W_{1,J},W_{3,J},V,Z)\nonumber\\
    & = & H(X|J,W_{1,J},W_{3,J},V_{J},Z) \nonumber \\
    & - & H(X|J,W_{1,J},W_{2,J},W_{3,J},W_{4,J},V_{J},Z)\nonumber\\
    &= & \frac{1}{N}\sum_{i=1}^N H(X_i|i,W_{1,i},W_{3,i},V_{i},Z_i)\nonumber\\
    & - &\frac{1}{N}\sum_{i=1}^N H(X_i|i,W_{1,i},W_{2,i},W_{3,i},W_{4,i},V_{i},Z_i)\nonumber\\
    & = & \frac{1}{N}\sum_{i=1}^N I(X_i;W_{2,i},W_{4,i}|W_{1,i},W_{3,i},V_{i},Z_i). \label{eq:conv20a}
\end{eqnarray}

It is left to prove that the cardinality of the auxiliary RVs
satisfies the third condition. This step of the proof extends the
converse proof of \cite{SMer04} and conceptually is very similar to
the above-detailed part of the converse proof of Theorem
\ref{theo_separation} which is related to reducing cardinality of
the alphabets of auxiliary RVs. The detailed proof of this part is
thus omitted and to complete the proof of the converse we merely
outline it. Here also we use the support lemma \cite{CK81} and
rewrite the relevant conditional mutual informations and the
distortion functions in a more convenient form for the use of this
lemma. Similarly as in \cite{SMer04}, we begin with the first term,
$I(X;W_1|Y)$, in the lower bound to $R_1$, using the Markov chain
$W_{1} \div X \div Y$:
\begin{eqnarray}
I(X;W_{1}|Y) & = & H(W_{1}|Y) - H(W_{1}|X,Y) \nonumber\\
            & = & H(W_{1}|Y) - H(W_{1}|X) \nonumber\\
            & = & H(W_{1}) - I(Y;W_{1}) - H(W_{1}) + I(X;W_{1})
            \nonumber \\
            & = & H(Y|W_{1}) - H(Y) - H(X|W_{1}) + H(X).
\end{eqnarray}
Next, we decompose the second term in the lower bound to $R_{1}$,
$I(X;V,W_{2}|W_{1},Z)$, into $I(X;V|W_{1},Z)$ and
$I(X;W_{2}|W_{1},V,Z)$, and for $I(X;V|W_{1},Z)$ we have due to the
Markov chain $(W_{1},V) \div X \div Z$:
\begin{eqnarray}
I(X;V|W_{1},Z) & = & H(X|W_{1},Z) - H(X|W_{1},V,Z) \nonumber\\
                & = & H(X|W_{1}) - I(X;Z|W_{1}) + I(X;Z|W_{1},V) - H(X|W_{1},V) \nonumber \\
                & = & H(X|W_{1}) - H(Z|W_{1}) + H(Z|W_{1},X) -
                H(X|W_{1},V) \nonumber %\\
\end{eqnarray}
\begin{eqnarray}
                & + & H(Z|W_{1},V) - H(Z|W_{1},V,X) \nonumber \\
                & = & H(X|W_{1}) - H(Z|W_{1}) + H(Z|W_{1},V) -
                H(X|W_{1},V).
\end{eqnarray}
Using the Markov chain $(W_{1},W_{2},V) \div X \div Z$ for
$I(X;W_{2}|W_{1},V,Z)$, we have:
\begin{eqnarray}
I(X;W_{2}|W_{1},V,Z) & = & H(X|W_{1},V) - H(Z|W_{1},V) \nonumber\\
                    & + & H(Z|W_{1},W_{2},V) - H(X|W_{1},W_{2},V).
\end{eqnarray}
Similarly, $I(X;W_{1},W_{3},V|Y)$ can be decomposed into
$I(X;W_{1}|Y)$, $I(X;V|W_{1},Y)$ and

\noindent $I(X;W_{3}|W_{1},V,Y)$, with two later terms, in turn,
expressed as
\begin{eqnarray}
I(X;V|W_{1},Y) = H(X|W_{1}) - H(Y|W_{1}) + H(Y|W_{1},V) -
                H(X|W_{1},V),
\end{eqnarray}
and
\begin{eqnarray}
I(X;W_{3}|W_{1},V,Y) & = & H(X|W_{1},V) - H(Y|W_{1},V) \nonumber\\
                    & + & H(Y|W_{1},W_{3},V) - H(X|W_{1},W_{3},V).
\end{eqnarray}
The second term in the lower bound to $R_{2}$ is
$I(X;W_{2},W_{4}|W_{1},V,W_{3},Z)$ and it can also be decomposed
into
\begin{eqnarray}
I(X;W_{2}|W_{1},W_{3},V,Z) & = & H(X|W_{1},W_{3},V) - H(Z|W_{1},W_{3},V) \nonumber\\
                    & + & H(Z|W_{1},W_{2},W_{3},V) - H(X|W_{1},W_{2},W_{3},V).
\end{eqnarray}
and
\begin{eqnarray}
I(X;W_{4}|W_{1},W_{2},W_{3},V,Z) & = & H(X|W_{1},W_{2},W_{3},V) - H(Z|W_{1},W_{2},W_{3},V) \nonumber\\
                    & + & H(Z|W_{1},W_{2},W_{3},W_{4},V) - H(X|W_{1},W_{2},W_{3},W_{4},V).
\end{eqnarray}
Thus, the lower bounds to $R_{1}$ and $R_{2}$ can be expressed as following:
\begin{eqnarray}
\label{eq:R1}
    I(X;W_{1}|Y) + I(X;V,W_{2}|W_{1},Z) & = & \big[H(X)-H(Y)\big]+\big[H(Y|W_{1})-H(Z|W_1)\big]\nonumber\\
                & + &\big[H(Z|W_{1},W_{2},V)-H(X|W_{1},W_{2},V)\big]
\end{eqnarray}
and
\begin{eqnarray}
\label{eq:R2}
    I(X;W_{1},W_{3},V|Y) + I(X;W_{2},W_{4}|W_{1},W_{3},V,Z)  &=& \big[H(X)-H(Y)\big]  \nonumber\\
    & + & \big[H(Y|W_{1},W_{3},V)-H(Z|W_1,W_{3},V)\big]\nonumber\\
    & + & \big[H(Z|W_{1},W_{2},W_{3},W_{4},V)\nonumber\\
    &-&H(X|W_{1},W_{2},W_{3},W_{4},V)\big].
\end{eqnarray}
Since $[H(X)-H(Y)]$ is a constant that depends only on the given statistics of the source and
SI $Y$, in order to preserve prescribed values of the above lower bounds,
it is sufficient to preserve the associated values of
$[H(Y|W_{1})-H(Z|W_1)] +[H(Z|W_{1},W_{2},V)-H(X|W_{1},W_{2},V)]$ and
$[H(Y|W_{1},W_{3},V)-H(Z|W_1,W_{3},V)]+[H(Z|W_{1},W_{2},W_{3},W_{4},V)-H(X|W_{1},W_{2},W_{3},W_{4},V)]$.

From here on the proof is essentially similar to the one provided
for Theorem \ref{theo_separation}: The support lemma is first used
to reduce the alphabet size of $W_{1}$, while preserving the values
of (\ref{eq:R1}) and (\ref{eq:R2}) and the distortions at both
stages. The alphabets of the remaining auxiliary RVs are kept intact
at this stage of the proof. There are $|\calX|-1$ functionals to be
defined that help to preserve the source distribution, $2$ more to
preserve (\ref{eq:R1}) and (\ref{eq:R2}) and $4$ more functionals to
preserve all the distortions at both stages. Thus, it is easy to
show that it is possible to find auxiliary RV $W_1$ which necessary
alphabet size is upper-bounded by $|\calX|+5$. Next, we reduce the
alphabet size of $V$, where now in addition to the values of the
lower bounds and distortions $\Delta_{z,1}$, $\Delta_{y,2}$ and
$\Delta_{z,2}$, it is desired to preserve the joint distribution
$(X,W_{1})$. There are $|\calX||\calW_{1}|-1 +2+ 3$ constraints
imposed on $V$ and thus its alphabet size is upper-bounded by
$|\calX|(|\calX|+5)+4$. In a similar manner, the reduction of the
alphabet cardinality is further performed for $W_{2}$, $W_{3}$ and
$W_{4}$ where at each stage, the support lemma is applied in so that
the statistics of the source and all already ``reduced" RVs are
maintained as well as lower bounds to the relevant rates and
distortions.

\subsection{Inner Bound}
\label{InnerBound-First}

%The achievability scheme is based on hierarchical coding and is
%heavily based on the assumption of degradedness of SI.
\subsubsection{Code-book generation} First, randomly generate,
according to $P_{W_{1}}(\cdot)$, a codebook $\calC_{w_{1}}$ of
$2^{[N(I(X;W_{1})+\epsilon_{1}+\delta)]}$ independent codewords
$\{\bw_{1,i}\}$ of length $N$, where the coordinates are also
generated i.i.d. Then, partition the codewords into
$2^{[N(I(X;W_{1}|Y)+\epsilon_{2}+\delta)]}$ bins ($\epsilon_{2} >
\epsilon_{1}$).

Next, for each $\{\bw_{1,i}\}$, randomly generate a codebook
$\calC_{v}(\bw_{1,i})$ consisting of \\
$2^{[N(I(X;V|W_{1})+\epsilon_{v}+\delta)]}$ codewords
$\{\bv_{i,j}\}$, where the generation of each coordinate is
according to $P_{V|W_{1}}(\cdot)$ and partition this codebook into
$2^{[N(I(X;V|W_{1},Z)+\epsilon_{v'}+\delta)]}$ bins,
$\calC_{v}(\bw_{1,i})$, ($\epsilon_{v'}
> \epsilon_{v}$). Each bin in the codebook of $\{\bv_{i,j}\}$ contains
a little less than $2^{[N(I(Z;V|W_{1}))]}$ codewords. Partition each
such bin into sub-bins, $\calC_{v}^{b}(\bw_{1,i})$, each of a size
of a little less than $2^{[N(I(Y;V|W_{1}))]}$. There are about
$2^{[N(I(Z;V|W_{1})-I(Y;V|W_{1}))]}$ such sub-bins.

For each pair $\{\bw_{1,i},\bv_{i,j}\}$ randomly generate a codebook
$\calC_{w_{2}}(\bw_{1,i},\bv_{i,j})$ consisting of\\
$2^{[N(I(X;W_{2}|W_{1},V)+\epsilon_{3}+\delta)]}$ codewords
$\{\bw_{2,i,j,k}\}$, where the generation of each coordinate is
according to $P_{W_{2}|W_{1},V}(\cdot)$ and partition
$\calC_{w_{2}}(\bw_{1,i},\bv_{j})$ into
$2^{[N(I(X;W_{2}|W_{1},V,Z)+\epsilon_{4}+\delta)]}$ bins
($\epsilon_{4}
> \epsilon_{3}$).

Now, randomly generate for each pair $(\bw_{1,i},\bv_{i,j})$ a
codebook $\calC_{w_{3}}(\bw_{1,i},\bv_{i,j})$ of\\
$2^{[N(I(X;W_{3}|W_{1},V)+\epsilon_{5}+\delta)]}$ codewords
$\{\bw_{3,i,j,l}\}$ according to $P_{W_{3}|W_{1},V}(\cdot)$ and
partition $\calC_{w_{3}}(\bw_{1,i},\bv_{j})$ into
$2^{[N(I(X;W_{3}|W_{1},V,Y)+\epsilon_{6}+\delta)]}$ bins
($\epsilon_{6}>\epsilon_{5}$).

Finally, for each quadruplet
$\{\bw_{1,i},\bw_{2,i,j,k},\bw_{3,i,j,l},\bv_{i,j}\}$, randomly
generate a codebook
\\$\calC_{w_{4}}(\bw_{1,i},\bv_{i,j},\bw_{2,i,j,k},\bw_{3,i,j,l})$ of
$2^{[N(I(X;W_{4}|W_{1},W_{2},W_{3},V)+\epsilon_{7}+\delta)]}$
codewords $\{\bw_{4,i,j,k,l,m}\}$ according to
$P_{W_{4}|W_{1},W_{2},W_{3},V}(\cdot)$ and partition it into
$2^{[N(I(X;W_{4}|W_{1},W_{2},W_{3},V,Z)+\epsilon_{8}+\delta)]}$ bins
($\epsilon_{8}>\epsilon_{7}$).

\begin{figure}[!htbp]
 \begin{center}
  \resizebox{0.6\textwidth}{!}
     {\includegraphics{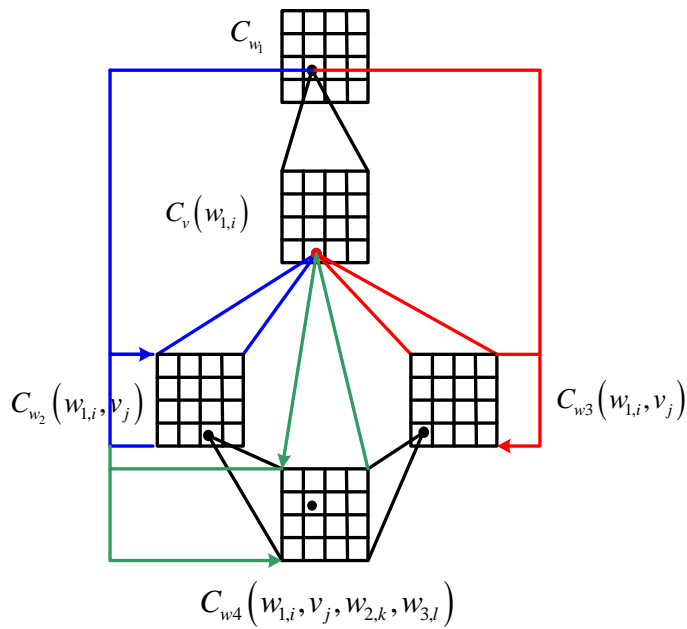}}
 \end{center}
\caption{Achievability Scheme - Code Generation.} \label{FAC}
\end{figure}

For clarity of exposition, the generation of codebooks is
demonstrated in Fig. \ref{FAC}.

\subsubsection{Encoding} Given a source sequence $\bx$,
the encoder seeks a vector in $\calC_{w_{1}}$ such that $\bx$ and
$\bw_{1,i}$ are jointly typical. If such $\bw_{1,i}$ is found, in
$\calC_{v}(\bw_{1,i})$, the encoder seeks a vector $\bv_{i,j}$ such
that the source sequences $\bx$ and $\bw_{1,i}$ will be jointly
typical with it. The encoder proceeds this way, seeking
$\bw_{2,i,j,k}$ in $\calC_{w_{2}}(\bw_{1,i},\bv_{i,j})$ so that
$(\bx,\bw_{1,i},\bv_{i,j},\bw_{2,i,j,k})$ are jointly typical. The
encoder then seeks in $\calC_{w_{3}}(\bw_{1,i},\bv_{i,j})$ a
codeword $\bw_{3,i,j,l}$ so that
$(\bx,\bw_{1,i},\bv_{i,j},\bw_{3,i,j,l})$ are jointly typical. Due
to the Markov chain $W_{2}\div(X,W_{1},V)\div W_{3}$, had the
encoder managed to find such sequences,
$(\bx,\bw_{1,i},\bv_{i,j},\bw_{2,i,j,k},\bw_{3,i,j,l})$ will be
jointly typical with high probability.

If the encoder found jointly typical sequences
$(\bx,\bw_{1,i},\bv_{i,j},\bw_{2,i,j,k},\bw_{3,i,j,l})$, it seeks in\\
$\calC_{w_{4}}(\bw_{1,i},\bv_{i,j},\bw_{2,i,j,k},\bw_{3,i,j,l})$ a
sequence $\bw_{4,i,j,k,l,m}$ which will be jointly typical with all
the above-mentioned sequences. If at any stage of its search the
encoder fails to find a ``good sequence", it declares an error. As
is shown in the sequel, the probability of such an event is very
low, due to the typicality properties of the scheme. Otherwise,
i.e., if all the jointly typical sequences are found, the encoder
acts as follows: At the first stage, it conveys to the decoders a
single transmission consisting of the following concatenated
indexes: the index $B_{1}$ of the bin to which $\bw_{1,i}$ belongs,
of length of about $NI(X;W_{1}|Y)$ bits; the index $B_{2}$ of
$\calC_{v}(\bw_{1,i})$, s.t., $\bv_{i,j}\in \calC_{v}(\bw_{1,i})$,
which can be described by about $NI(X;V|W_{1},Z)$ bits and the index
$B_{3}$ of the bin to which $\bw_{2,i,j,k}$ belongs, which requires
about $NI(X;W_{2}|W_{1},V,Z)$ bits. At the refinement stage, it
transmits the index $B^{*}_{4}$ of $\calC^{b}_{v}(\bw_{1,i})$ to
which $\bv_{i,j}$ belongs \textit{within} $\calC_{v}(\bw_{1,i})$
(previously described by $B_{2}$), which requires about
$N[I(Z;V|W_{1})-I(Y;V|W_{1})]$ bits, concatenated with the indexes
$B_{5}$ and $B_{6}$ of the bins containing $\bw_{3,i,j,l}$ and
$\bw_{4,i,j,k,l,m}$, in $C_{w3}(\bw_{1,i},\bv_{i,j})$ and
$C_{w4}(\bw_{1,i},\bv_{i,j},\bw_{2,i,j,k},\bw_{3,i,j,l})$, of about
$NI(X;W_{3}|W_{1},V,Y)$ and $NI(X;W_{4}|W_{1},W_{2},W_{3},V,Z)$
bits, respectively. The transmission rates at both stages are as
defined by $\calR^{*}(\bD)_{nc}$ up to $\{\epsilon_{i}\}$.

\subsubsection{Decoding}

\textbf{First stage:} The first decoder accesses
$(B_{1},B_{2},B_{3})$, but performs W-Z decoding procedure with
respect to $B_{1}$ only. Specifically, in $\calC_{w_{1}}$, in the
bin indexed by $B_{1}$, the decoder seeks a unique sequence
$\bw_{1,i}$ that was chosen by the encoder. Due to the
Markov chain $W_{1}\div X \div Y$, as the block-length becomes
infinitely large, the decoder will find with probability tending to
$1$ the correct sequence $\bw_{1,i}$. Since
in each bin in $\calC_{w_{1}}$ there are less than $2^{NI(Y;W_{1})}$
codewords, and these codewords were generated i.i.d, the probability
of existing at the bin indexed by $B_{1}$ of another codeword
jointly typical with $\bY$ vanishes as $N \rightarrow \infty$.

The second decoder uses three indexes $(B_{1},B_{2},B_{3})$ to
retrieve all three codewords chosen by the encoder. Specifically,
it retrieves $\bw_{1,i}$ similarly as Y-decoder does, since, as it
has access to a more informative SI, it can do whatever the
Y-decoder can do. Afterwards, it retrieves correctly
$\bv_{i,j}\in\calC_{v}(\bw_{1,i})$ in the bin indexed by $B_{2}$,
which is possible due to the Markov chain $(V,W_{1})\div X \div Z$.
The Z-decoder does not find in bin indexed by $B_{2}$ other
codewords which are jointly typical with $\bz$ since there are less
than $2^{NI(Z;V|W_{1})}$ codewords in that bin. Finally, following
similar considerations, after retrieving $(\bw_{1,i},\bv_{i,j})$,
the Z-decoder retrieves correctly $\bw_{2,i,j,k}\in
\calC_{w_{2}}(\bw_{1,i},\bv_{i,j})$ in the bin indexed by $B_{3}$.

%After each of the decoders has found correct codewords, it performs reconstruction of the source sequence $\bx$. Due
%to the typicality properties of the scheme, i.e., $X\div (W_{1},Y)
%\div \hat{X}(W_{1},Y)$ and $X\div (W_{1},W_{2},V,Z) \div
%\tilde{X}(W_{1},W_{2},V,Z)$, the distortion constraints are
%satisfied at both decoders.

\textbf{Second Stage:} Note that after the first transmission
Y-decoder is able to find all codewords $\bv$ which are jointly
typical with $\by$ in the bin indexed by $B_{2}$ in the codebook
$\calC_{v}^{b}(\bw_{1,i})$. This is due to the Markov chain
$(W_{1},V) \div X \div Y$. But, it cannot reveal which of these
codewords was chosen by the encoder, as there are more than
$2^{NI(Y;V|W_{1})}$ such codewords (there are a bit less then
$2^{NI(Z;V|W_{1})}$ such codewords, as is required by the W-Z coding
designed for Z-decoder). When the Y-decoder receives the index
$B^{*}_{4}$ of $\calC^{b}_{v}(\bw_{1,i})$, since $\bv_{i,j} \in
\calC^{b}_{v}(\bw_{1,i}) \subseteq \calC_{v}(\bw_{1,i})$, it
searches $\bv_{i,j}$ among a group of codewords of a size less than
$2^{NI(Y;V|W_{1})}$ codewords, and thus, it is is able to retrieve
$\bv_{i,j}$ correctly by the W-Z decoding argument.
%
%Conveying the index $B^{*}_{4}$ of $\calC^{b}_{v}(\bw_{1,i})$ helps
%to focus the attention of the Y-decoder on a smaller group of
%codewords. This is because, as defined previously $\bv_{i,j} \in
%\calC^{b}_{v}(\bw_{1,i}) \subseteq \calC_{v}(\bw_{1,i})$. The size
%of $\calC^{b}_{v}(\bw_{1,i})$ is less than $2^{NI(Y;V|W_{1})}$
%codewords, and thus, Y-decoder is able to retrieve correctly
%$\bv_{i,j}$.
%
After Y-decoder has found $\bv_{i,j}$, it performs W-Z decoding of
the codeword $\bw_{3,i,j,l} \in \calC_{w_{3}}(\bw_{1,i},\bv_{i,j})$
according to the bin-index $B_{5}$ and $(\bw_{1,i},\bv_{i,j})$. It
now improves the reconstruction of the source sequence with an aid
of the triplet $(\bw_{1,i},\bv_{i,j},\bw_{3,i,j,l})$, which is
possible within the defined distortion due to the typicality
properties of the scheme.

The Z-decoder, which after the first step has retrieved correctly
(with probability tending to 1, as $N\rightarrow \infty$) the
sequences $(\bw_{1,i},\bv_{i,j},\bw_{2,i,j,k})$, makes no use of
index $B_{4}^{*}$, as it serves Y-decoder only. The Z-decoder uses
its knowledge of $(\bw_{1,i},\bv_{i,j})$ as well as the fact that
its SI is more informative to decode correctly $\bw_{3,i,j,l}$ in
the bin of $\calC_{w_{3}}(\bw_{1,i},\bv_{i,j})$ indexed by $B_{5}$.
Finally, it uses all the codewords it managed to find thus far to
perform conditional W-Z decoding and to find the correct codeword
$\bw_{4,i,j,k,l,m}$ according to the index $B_{6}$ of a bin in
$\calC_{w_{4}}(\bw_{1,i},\bv_{i,j},\bw_{2,i,j,k},\bw_{3,i,j,l})$.

At each stage, after each of the decoders has found correct
codewords, it performs reconstruction of the source sequence $\bx$.
Due to the typicality properties of the scheme, i.e., $X\div
(W_{1},Y) \div \hat{X}$, $X\div (W_{1},W_{2},V,Z) \div \tilde{X}$,
$X\div (W_{1},W_{3},V,Y) \div \check{X}$ and $X\div
(W_{1},W_{2},W_{3},W_{4},V,Z) \div \bar{X}$, the distortion
constraints are satisfied at both decoders.

\subsubsection{Analysis of Probability of Error}
%The probability of error in decoding of either of the decoders, can
%be written explicitly following the lines of the direct proofs in
%\cite{SMer04}. Similarly as in \cite{SMer04}, it can be shown that
%the probability of decoding error vanishes when the block-length is
%sufficiently large and for appropriate choices of
%$\{\epsilon_{i}\}$.
We now turn to the analysis of the error probability. For each $\bx$
and a particular choice of the code $\calC_{w_1}$ and related
choices of
$(\{\calC_{v}(\cdot),\calC_{w_2}(\cdot),\calC_{w_3}(\cdot),\calC_{w_4}(\cdot)\})$,
the possible causes for error message are:

\begin{enumerate}
  \item {$\bx \notin \Tgx$. Let the probability of this event be defined as $P_{e_{1}}$.
  }
  \item {$\bx \in \Tgx$, but in the codebook $\calC_{w_{1}}$ $\not\exists \bw_{1,i}$ s.t. $(\bx,\bw_{1,i}) \in \Tgxwa$.
    Let the probability of this event be defined as $P_{e_{2}}$.
  }
  \item {$\bx \in \Tgx$, and the codebook $\calC_{w_{1}}$ contains $\bw_{1,i}$ s.t. $(\bx,\bw_{1,i})
    \in \Tgxwa$, but $\not\exists \bv_{i,j} \in \calC_{v}(\bw_{1,i})$ s.t. $(\bx,\bw_{1,i},\bv_{i,j}) \in \Tgxwav$.
    Let the probability of this event be defined as $P_{e_{3}}$.
  }
  \item {$\bx \in \Tgx$, the codebook $\calC_{w_{1}}$ contains $\bw_{1,i}$ s.t. $(\bx,\bw_{1,i})
    \in \Tgxwa$, and also the codebook $\calC_{v}(\bw_{1,i})$
    contains $\bv_{i,j}$ s.t. $(\bx,\bw_{1,i},\bv_{i,j}) \in
    \Tgxwav$, but $\not\exists \bw_{2,i,j,k} \in \calC_{w_{2}}(\bw_{1,i},\bv_{i,j})$ s.t. $(\bx,\bw_{1,i},\bv_{i,j},\bw_{2,i,j,k}) \in \Tgxwavwb$.
    Let the probability of this event be defined as $P_{e_{4}}$.
  }
  \item {$\bx \in \Tgx$, the codebook $\calC_{w_{1}}$ contains $\bw_{1,i}$ s.t. $(\bx,\bw_{1,i})
    \in \Tgxwa$, the codebook $\calC_{v}(\bw_{1,i})$
    contains $\bv_{i,j}$ s.t. $(\bx,\bw_{1,i},\bv_{i,j}) \in
    \Tgxwav$, but $\not\exists \bw_{3,i,j,l} \in \calC_{w_{3}}(\bw_{1,i},\bv_{i,j})$ s.t. $(\bx,\bw_{1,i},\bv_{i,j},\bw_{3,i,j,l}) \in \Tgxwavwc$.
    Let the probability of this event be defined as $P_{e_{5}}$.
  }
  \item {$\bx \in \Tgx$, the codebook $\calC_{w_{1}}$ contains $\bw_{1,i}$ s.t. $(\bx,\bw_{1,i})
    \in \Tgxwa$, the codebook $\calC_{v}(\bw_{1,i})$
    contains $\bv_{i,j}$ s.t. $(\bx,\bw_{1,i},\bv_{i,j}) \in
    \Tgxwav$, and the codebooks $\calC_{w_{2}}(\bw_{1,i},\bv_{i,j})$ and $\calC_{w_{3}}(\bw_{1,i},\bv_{i,j})$
    contain $\bw_{2,i,j,k}$ s.t. $(\bx,\bw_{1,i},\bv_{i,j},\bw_{2,i,j,k}) \in \Tgxwavwb$ and
    $\bw_{3,i,j,m}$ s.t. $(\bx,\bw_{1,i},\bv_{i,j},\bw_{3,i,j,m}) \in
    \Tgxwavwc$, respectively, but $(\bx,\bw_{1,i},\bv_{i,j},\bw_{2,i,j,k},\bw_{3,i,j,l}) \notin \Tgxwavwbwc$.
    Let the probability of this event be defined as $P_{e_{6}}$.
  }
  \item {$\bx \in \Tgx$, the codebook $\calC_{w_{1}}$ contains $\bw_{1,i}$ s.t. $(\bx,\bw_{1,i})
    \in \Tgxwa$, the codebook $\calC_{v}(\bw_{1,i})$
    contains $\bv_{i,j}$ s.t. $(\bx,\bw_{1,i},\bv_{i,j}) \in
    \Tgxwav$, and the codebooks $\calC_{w_{2}}(\bw_{1,i},\bv_{i,j})$ and $\calC_{w_{3}}(\bw_{1,i},\bv_{i,j})$
    contain $\bw_{2,i,j,k}$ s.t. $(\bx,\bw_{1,i},\bv_{i,j},\bw_{2,i,j,k}) \in \Tgxwavwb$ and
    $\bw_{3,i,j,m}$ s.t. $(\bx,\bw_{1,i},\bv_{i,j},\bw_{3,i,j,m}) \in
    \Tgxwavwc$, respectively, and also $(\bx,\bw_{1,i},\bv_{i,j},\bw_{2,i,j,k},\bw_{3,i,j,l}) \in
    \Tgxwavwbwc$, but$\not\exists \bw_{4,i,j,l,k,m} \in \calC_{w_{4}}(\bw_{1,i},\bv_{i,j},\bw_{2,i,j,k},\bw_{3,i,j,l})$
    s.t. \\$(\bx,\bw_{1,i},\bv_{i,j},\bw_{2,i,j,k},\bw_{3,i,j,l},\bw_{4,i,j,k,l,m}) \in \Tgxwavwbwc$.
    Let the probability of this event be defined as $P_{e_{7}}$.
  }\end{enumerate}

Note that if none of those events occur, then, for the sufficiently
large $N$, by the Markov Lemma \cite[pp.\ 436, Lemma 14.8.1]{CT91}
applied twice, the following is satisfied: with high probability
$(\bX,\bY,\mathbf{\hat{X}})$ are jointly typical and
$(\bX,\bZ,\mathbf{\tilde{X}})$ are jointly typical at both stages.
\begin{enumerate}
\item{The first application of the Markov Lemma occurs due to
the Markov chain $ (Y,Z)\div X\div(W_{1},V,W_{2},W_{3},W_{4})$: Note
that by the way of creation, $\bX$, $\bY$ and $\bZ$ are jointly
typical with high probability and also, with high probability, RV's
$(\bW_1,\bW_2,\bW_{3},\bW_{4},\bV)$ and $\bX$ are jointly typical.
Therefore, by the Markov Lemma, all the sequences $\bX$, $\bY$,
$\bZ$, $\bW_1$, $\bW_2$, $\bW_{3}$, $\bW_{4}$ and $\bV$ are also
jointly typical with high probability. And so, SIs are jointly
typical with the auxiliary RV's at both stages of communication.}
\item{ Also, note that due
to the fact that the source is memoryless and by the way of creation
of the reconstructions, the following Markov chains hold at the
first stage: $\bX\div (\bY,\bW_1) \div \mathbf{\hat{X}}$ and
$\bX\div (\bZ,\bW_1,\bV,\bW_{2}) \div \mathbf{\tilde{X}}$.
Similarly, at the second stage, $\bX\div (\bY,\bW_1,\bV,\bW_{3})
\div \mathbf{\hat{X}}$ and $\bX\div
(\bZ,\bW_1,\bV,\bW_{2},\bW_{3},\bW_{4}) \div \mathbf{\tilde{X}}$. By
the second application of the Markov Lemma, we obtain that with high
probability $\bX$ is jointly typical with $\mathbf{\hat{X}}$ and
$\mathbf{\tilde{X}}$ at both stages. The probability that one or
more of the above typicality relations do not hold vanishes as $N$
becomes infinitely large. The joint typicality of
$(\bX,\mathbf{\hat{X}})$ and $(\bX,\mathbf{\tilde{X}})$ imposes that
the distortion constraints (\ref{theo_1c3ll})- (\ref{theo_1d3a}) are
satisfied when $N$ is large enough (see \cite[Section 6]{SMer04} for
explicit derivations). }
\end{enumerate}

It remains to show that the probability of sending an error message
vanishes when $N$ is large enough. The average probability of error
$P_{e}$ is bounded by
\begin {eqnarray}
\label{Pe}
    P_{e} \leq
    P_{e_{1}}+P_{e_{2}}+P_{e_{3}}+P_{e_{4}}+P_{e_{5}}+P_{e_{6}}+P_{e_{7}}.
\end {eqnarray}

\noindent The fact that $P_{e_{1}}\rightarrow 0$ follows from the
properties of typical sequences \cite{CT91}. As for $P_{e_{2}}$, we
have:
\begin {equation}
\label{P_e2a}
    P_{e_{2}} \eql \prod_{k=1}^{|\calC_{w_{1}}|}\Pr\left\{\left(\bx,\bW_{1,k}\right) \notin \Tgxwa\right\}.
\end {equation}
Now, for every $k$:
\begin {eqnarray}
\label{P_e2i}
    \Pr\left\{\left(\bx,\bW_{1,k}\right) \notin \Tgxwa\right\} & = & 1 -
        \Pr\left\{\left(\bx,\bW_{1,k}\right) \in \Tgxwa\right\} \\
    & = & 1 - \frac{|\Tgxwa|}{|\Tgwa||\Tgx|} \nonumber\\
    & \leq &1 - 2^{-N[I(X;W_{1})+\epsilon_{1}]} \nonumber,
\end {eqnarray}
where the last equation follows from the size of typical sequences
as are given in \cite{CT91}. Substitution of (\ref{P_e2i}) into
(\ref{P_e2a}) and application of the well-known inequality
$(1-v)^{N}\leq\exp(-vN)$, provides us with the following upper-bound
for $N\rightarrow\infty$:
\begin {equation}
    P_{e_{2}} \leq \Big[1 - 2^{-N[I(X;W_{1})+\epsilon_{1}]}\Big]^{|\calC_{w_{1}}|} \leq
        \exp\left\{-|\calC_{w_{1}}|\cdot2^{-N[I(X;W_1)+\epsilon_{1}]}\right\} \rightarrow 0,
\end {equation}
double-exponentially rapidly since $|\calC_{w_{1}}| =
I(X;W_{1})+\epsilon_{1} + \delta$.

\indent To estimate $P_{e_{3}}$, we repeat the technique of the
previous step:
\begin {equation}
\label{P_e3a}
    P_{e_{3}} \eql \prod_{j=1}^{|\calC_{v}|}\Pr\left\{\left(\bx,\bw_{1,i},\bV_{i,j}\right) \notin \Tgxwav\right\}.
\end {equation}
Again, by the property of the typical sequences, for every $j$:
\begin {eqnarray}
\label{P_e3Cu}
    \Pr\left\{\left(\bx,\bw_{1},\bV_{i,j}\right) \notin \Tgxwav\right\}
    \leq 1 - 2^{-N[I(X;V|W_{1})+\epsilon_{2}]},
\end {eqnarray}
and therefore, substitution of (\ref{P_e3Cu}) into (\ref{P_e3a})
gives
\begin {equation}
    P_{e_{3}} \leq \Big[1 - 2^{-N[I(X;V|W_{1})+\epsilon_{2}]}\Big]^{|\calC_{v}|} \leq
        \exp\left\{-|\calC_{v}|\cdot2^{-N[I(X;V|W_{1})+\epsilon_{2}]}\right\} \rightarrow 0,
\end {equation}
double-exponentially rapidly since $|\calC_{v}| =
I(X;V|W_{1})+\epsilon_{2} + \delta$.

\indent To estimate $P_{e_{4}}$, the technique of the previous step
is again repeated:
\begin {equation}
\label{P_e4a}
    P_{e_{4}} \eql \prod_{k=1}^{|\calC_{w_{2}}|}\Pr\left\{\left(\bx,\bw_{1,i},\bv_{i,j},\bW_{2,i,j,k}\right) \notin \Tgxwavwb\right\}.
\end {equation}
Still, by the property of the typical sequences, for every $k$:
\begin {eqnarray}
\label{P_e4Cu}
    \Pr\left\{\left(\bx,\bw_{1,i},\bv_{i,j},\bW_{2,i,j,k}\right) \notin \Tgxwavwb\right\}
    \leq 1 - 2^{-N[I(X;W_{2}|W_{1},V)+\epsilon_{3}]},
\end {eqnarray}
and therefore, substitution of (\ref{P_e4Cu}) into (\ref{P_e4a})
gives
\begin {equation}
    P_{e_{4}} \leq \Big[1 - 2^{-N[I(X;W_{2}|W_{1},V)+\epsilon_{3}]}\Big]^{|\calC_{w_{2}}|} \leq
        \exp\left\{-|\calC_{w_{2}}|\cdot2^{-N[I(X;W_{2}|W_{1},V)+\epsilon_{3}]}\right\} \rightarrow 0,
\end {equation}
double-exponentially rapidly since $|\calC_{w_{2}}| =
I(X;W_{2}|W_{1},V)+\epsilon_{3} + \delta$.

Similarly as in the previous step we show that $P_{e_{5}}$ and
$P_{e_{7}}$ vanishes as well when $N$ is large enough, using the
fact that $|\calC_{w_{3}}| = I(X;W_{3}|W_{1},V)+\epsilon_{6} +
\delta$ and $|\calC_{w_{4}}| =
I(X;W_{4}|W_{1},W_{2},W_{3},V)+\epsilon_{7} + \delta$, respectively.

The proof for $P_{e_{6}}$ is different and it uses the Markov lemma
\cite[pp.\ 436, Lemma 14.8.1]{CT91}. In the previous steps we show
that the probability that the quadruples $(\bX,\bW_{1},\bV,\bW_{2})$
and $(\bX,\bW_{1},\bV,\bW_{3})$ are jointly typical with high
probability. Now, due to the Marlov lemma applied to the Markov
chain $W_{2} \div (X,W_{1},V) \div W_{3}$, the probability that
$(\bX,\bW_{1},\bV,\bW_{2},\bW_{3})$ are not typical tends to zero
with $N$ approaching infinity. Therefore, $P_{e_{6}}\rightarrow 0$
when $N \rightarrow \infty$.

\indent Since $P_{e_{s}}\rightarrow 0$ for $s\in[1,7]$, their sum
tends to zero as well, implying that there exist at least one choice
of a codebook $\calC_{w_1}$ and related choices of sets $\{\calC_{v}\}$,
$\{\calC_{w_2}\}$, $\{\calC_{w_3}\}$, $\{\calC_{w_4}\}$ that give rise to the
reliable source reconstruction at both stages with communication
rates $R_{1}$ and $R_{2}$.

\end{document}